\documentclass[epj]{svjour}
\usepackage{graphicx,bm,amsmath,macros,amssymb,color,empheq,nicefrac,multirow,cite,psfrag}

\title{Stretching dynamics of semiflexible polymers} \author{B.
  Obermayer\inst{1}\thanks{\email{obermayer@physik.lmu.de}} \and O.
  Hallatschek\inst{2} \and E. Frey\inst{1} \and K. Kroy\inst{3}}
\institute{ \inst{1} Arnold Sommerfeld Center and Center for
  NanoScience,
  LMU M\"unchen, Theresienstr. 37, 80333 M\"unchen, Germany\\
  \inst{2} Lyman Laboratory of Physics, Harvard University, Cambridge,
  MA 02138, USA\\
  \inst{3} Institut f\"ur Theoretische Physik, Universit\"at Leipzig
  Postfach 100920, 04009 Leipzig, Germany} 

\abstract{We analyze the nonequilibrium dynamics of single
  inextensible semiflexible biopolymers as stretching forces are
  applied at the ends. Based on different (contradicting) heuristic
  arguments, various scaling laws have been proposed for the
  propagation speed of the backbone tension which is induced in
  response to stretching.  Here, we employ a newly developed unified
  theory to systematically substantiate, restrict, and extend these
  approaches.  Introducing the practically relevant scenario of a
  chain equilibrated under some prestretching force $\fpre$ that is
  suddenly exposed to a different external force $\fext$ at the ends,
  we give a concise physical explanation of the underlying relaxation
  processes by means of an intuitive blob picture. We discuss the
  corresponding intermediate asymptotics, derive results for
  experimentally relevant observables, and support our conclusions by
  numerical solutions of the coarse-grained equations of motion for
  the tension.}

\begin{document}
\maketitle

The multifaceted viscoelastic properties of complex cellular
structures such as the cytoskeleton have been intensively studied on
very different levels of complexity~\cite{bausch-kroy:06}. Concerning
their nature as networks of semiflexible polymers, it has recently
become apparent that \emph{prestressed} networks provide a realistic
model for the mechanical properties of living
cells~\cite{gardel-etal:06}. Also, it has been found that the
\emph{dynamics} of semiflexible polymers and their networks is quite
drastically altered by
prestress~\cite{granek:97,rosenblatt-etal:06,mizuno-tardin-schmidt-mackintosh:07}.
On the level of single molecules, the scenario of a single
semiflexible chain prestretched with a force $\fpre$ that is suddenly
exposed to a different stretching force $\fext$ at the ends may serve
to explain quite generically the influence of prestress on
single-polymer stretching and relaxation dynamics. Since the first
force-extension measurements on DNA, which established the wormlike
chain (WLC) model as a very good theoretical description for
semiflexible polymers~\cite{bustamante-bryant-smith:03}, technological
advances towards significantly improved time- and
force-resolution~\cite{meiners-quake:00,lumma-etal:03} have made
measurements on such more involved experimental setups possible.
Theoretically, there exists a number of different approaches to the
nonequilibrium dynamics of inextensible
WLCs~\cite{goldstein-langer:95,lee-thirumalai:04,bohbot_raviv-etal:04,liverpool:05},
most of them based on the concept of backbone tension, which is the
force ``holding the monomers together''. Its key importance to the
longitudinal dynamics has been recognized some time
ago~\cite{seifert-wintz-nelson:96}, but so far its influence has
mainly been treated on the level of heuristic scaling
arguments~\cite{seifert-wintz-nelson:96,ajdari-juelicher-maggs:97,everaers-juelicher-ajdari-maggs:99,brochard-buguin-degennes:99}.
While such reasoning based on the WLC model has been used to analyze
DNA relaxation
experiments~\cite{manneville-etal:96,maier-seifert-raedler:02}, a
systematic theoretical description of propagation and relaxation of
backbone tension has only recently been
developed~\cite{hallatschek-frey-kroy:05}.

In the following, we consider an inextensible weakly-bending WLC in a
simple solvent that is equilibrated under some tension $\fpre$. At
time $t=0$, the chain is suddenly exposed to an external stretching
force $\fext$ applied at its ends, the more interesting case being
$\fext > \fpre$.  Our objective is to identify the dominant processes
involved in the subsequent nonequilibrium dynamics, and, specifically,
to determine how fast the contour stretches. The inextensibility
prevents any contour stretching in the bulk unless the end regions
have been pulled apart, which is however limited by longitudinal
Stokes friction. The conclusion is that initially the external
stretching force penetrates the contour only within a boundary layer
of size $\lpar(t)$, see fig.~\ref{fig:blob-picture}.  This comprises
the relevant short-time effect and motivates the analysis presented in
the remainder of the paper. After introducing model and equations of
motion in sec.~\ref{sec:eom}, we discuss in
sec.~\ref{sec:blob-picture} the stretching dynamics qualitatively by
means of an intuitive ``blob'' picture inspired by the analogon for
flexible polymers~\cite{degennes-pincus-velasco-brochard:76}. Hereby
we regard only the special case $\fext\gg\fpre$ since it captures all
prominent features of the stretching dynamics.  We obtain crossover
scaling laws for $\lpar(t)$ that are compared to literature
results~\cite{seifert-wintz-nelson:96,ajdari-juelicher-maggs:97,everaers-juelicher-ajdari-maggs:99,brochard-buguin-degennes:99}.
In sec.~\ref{sec:tension-profiles}, we explicitly calculate tension
profiles by solving the equations of motion for the asymptotic regimes
of sec.~\ref{sec:blob-picture}, and point out some inconsistencies of
previous
approaches~\cite{seifert-wintz-nelson:96,brochard-buguin-degennes:99}.
In the following sec.~\ref{sec:numerics} we present an algorithm to
solve the equations of motion for the tension in intermediate and
non-asymptotic regimes. Analytical and numerical results for
experimentally relevant observables such as the change in the
polymer's projected length are presented in the final
sec.~\ref{sec:deltar}, where also some experimental implications are
shortly discussed. A more systematic derivation of the asymptotic
equations of motion and their solutions, as well as details of the
numerical approach are shifted to the appendices.

\section{\label{sec:eom}Equations of motion}

Conformations of wormlike chains are described as continuous space
curves $\bvec r (s,t)$ of total contour length $L$. In the
WLC-Hamiltonian~\cite{saito-takahashi-yunoki:67}, bending energy is
proportional to the square of the local curvature.\footnote{Note that
  this harmonic dependence has lately been under some debate
  concerning DNA bending angle distributions on short length
  scales~\cite{wiggins-etal:06,mazur:07}.}  We set the proportionality
constant (the bending rigidity) $\kappa\equiv 1$.  Below, we will also
set the friction coefficient (per length) $\zeta$ to unity. In these
units, tension is a length$^{-2}$ and time a length$^4$. The
persistence length of a free WLC in equilibrium is then given by
$\lp=(\kb T)^{-1}$. We account for the local inextensibility
$\abs{\bvec r'}\equiv\abs{\pd_s\bvec r}=1$ by introducing the backbone
tension $f(s,t)$ as a Lagrange multiplier
function~\cite{goldstein-langer:95}.  In the limit of small transverse
displacements $\rperp$ from a straight line, the corresponding
\emph{weakly-bending} Hamiltonian reads
\begin{equation}\label{eq:hamiltonian}
  \mathcal{H} = \frac{1}{2}\int_0^L\!\!\td s [\rperp''^2 + f \rperp'^2].
\end{equation}
Comparing contributions from bending $\rperp^2/\lb^4$ and from tension
$f\rperp^2/\lb^2$ on the scaling level, we infer that on length scales
smaller than the \emph{blob size} $\lb\equiv
f^{-1/2}$~\cite{kierfeld-etal:04} (see fig.~\ref{fig:blob-picture}),
the conformation is dominated by bending forces, and only on larger
scales perturbed by the tension contributions. In equilibrium, each
blob carries a stored length of
$\lb^2/\lp$~\cite{saito-takahashi-yunoki:67}, which is simply the
thermal contraction compared to the straight conformation.  The
weakly-bending assumption requires $\lb\ll \lp$. This can be realized
both for short stiff polymers ($L\ll\lp$) and for strongly
prestretched flexible WLCs ($\fpre\gg \lp^{-2}$).

The overdamped motion of the transverse displacements is described by
the equation $\pd_t \rperp = -\delta \mathcal{H}/\delta \rperp + \bvec
\xi_\perp$, which captures the balance of viscous friction, bending
and tension forces resulting from the energy
[eq.~\eqref{eq:hamiltonian}], and stochastic noise $\bvec \xi_\perp$:
\begin{equation}\label{eq:eom-t}
  \pd_t \rperp = -\rperp''''+ (f\rperp')' + \bvec \xi_\perp.
\end{equation}
Introduced as Lagrange multiplier function, the tension $f(s,t)$ has
to be computed through the inextensibility constraint, which makes
eq.~\eqref{eq:eom-t} highly nonlinear. In spite of that, it can be
simplified quite significantly in the weakly-bending
limit~\cite{hallatschek-frey-kroy:05}: spatial variations in the
tension $f$ can be neglected.  It turns thus into a linear equation
that can be decomposed into normal modes,
\begin{equation}\label{eq:eom-t-modes}
  \pd_t \rperp = - q^4 \rperp- q^2 f \rperp + \bvec \xi_\perp,
\end{equation}
and solved using the response function
\begin{equation}\label{eq:chi-perp}
  \chi_\perp(q;t,t') = \e^{-q^2[q^2(t-t')+\int_{t'}^t\!f \td \tilde t]}\Theta(t-t').
\end{equation}
Here, $\Theta(t)$ is the Heaviside function. This allows to evaluate a
characteristic quantity of high relevance for our approach: the
\emph{stored length density} $\varrho\equiv\frac{1}{2}\ave{\rperp'^2}$
that measures the average amount of contour length stored in thermal
undulations:
\begin{equation}\label{eq:varrho-discrete}
  \varrho = \sum_q \left[\varrho_0(q)\chi^2_\perp(q;t,0)
    +\frac{2 q^2}{L\lp} \int_0^t\!\td t'\chi^2_\perp(q;t,t')\right],
\end{equation}
by evaluating the noise correlation
$\ave{\bvec\xi_{\perp,k}(t)\bvec\xi_{\perp,q}(t')}=4/(L \lp)
\delta_{k,q}\delta(t-t')$. The initial mode spectrum of a polymer
equilibrated under the force $\fpre$ follows via equipartition from
the Hamiltonian [eq.~\eqref{eq:hamiltonian}]:
\begin{equation}\label{eq:initial-spectrum}
  \varrho_0(q) = \frac{1}{L\lp (q^2 + \fpre)}.
\end{equation}

However, as discussed in the introduction and illustrated in
fig.~\ref{fig:blob-picture}, neither the tension $f$ nor the stored
length density $\varrho$ are in fact spatially constant. Compared to
the ``fast'' transverse undulations though, their spatial dependence
is rather slow and can thus be re-introduced within an adiabatic
approximation.  This convenient practice is rigorously justified by
means of a multiple scale analysis in
ref.~\cite{hallatschek-frey-kroy:07a}. Its main result is that the
curvature in the spatially slowly varying tension profile is given by
changes in the stored length density $\varrho(s,t)$, which inherits
its slow arc\-length dependence adiabatically from the tension
$f(s,t)$ via eq.~\eqref{eq:chi-perp}:
\begin{equation}\label{eq:eom-l}
  \pd_s^2 f(s,t) = -\pd_t \varrho(s,t).
\end{equation}
Integrating eq.~\eqref{eq:eom-l} over time yields
\begin{equation}\label{eq:eom-l-integrated}
  \pd_s^2 F(s,t) = - [\varrho(s,t)-\varrho(s,0)],
\end{equation}
where we have introduced the integrated tension
\begin{equation}\label{eq:integrated-tension}
  F(s,t)\equiv\int_0^t\!\td t'f(s,t'). 
\end{equation}
Using eq.~\eqref{eq:varrho-discrete} with a spatial dependence of
$\varrho(s,t)$ given through $f(s,t)$ in the exponent of
$\chi_\perp(q;t,t')$, and taking the continuum limit $L\to\infty$,
renders eq.~\eqref{eq:eom-l-integrated} in the explicit
form~\cite{hallatschek-frey-kroy:07a}
\begin{multline}\label{eq:pide}
  \pd_s^2 F(s,t) = \int_0^\infty\!\!\frac{\td q}{\pi\lp} \Bigg\lbrace
  \frac{1-\e^{-2 q^2[q^2 t + F(s,t)]}}{q^2+\fpre}  \\
  - 2 q^2\!\int_0^t\!\td t'\,\e^{-2 q^2 [q^2(t-t') +
    F(s,t)-F(s,t')]}\Bigg\rbrace.
\end{multline}
Initial and boundary conditions for our setup are
\begin{equation}\label{eq:pide-bc}
  F(s,t < 0)=\fpre t \quad\text{ and }  F(0,t)=F(L,t)=\fext t.
\end{equation}

For a detailed derivation of eq.~\eqref{eq:pide} we refer the
interested reader to ref.~\cite{hallatschek-frey-kroy:07a}.  Here we
only comment on a few important points. First, a proper decomposition
of eq.~\eqref{eq:eom-t} into its eigenmodes depends on the choice of
boundary conditions and yields, in general, eigenvalues different from
the simple $q^2[q^2 +f]$ used here.  Second, we are going to extend
eq.~\eqref{eq:pide}, which applies to the continuum limit
$L\to\infty$, to ``real'' polymers of finite length. Both issues are
resolved by recognizing that on short times the \emph{relevant} modes
have wavelength much shorter than $L$, that long wavelengths are
initially suppressed by prestretching with $\fpre$, and that therefore
the influence of the boundary conditions and the discreteness of the
mode spectrum is small. Below, we will discuss the relevance of these
effects for pertinent observables. Finally, the underlying scale
separation between small-scale ``fast'' transverse fluctuations and
large-scale ``slow'' tension dynamics can be readily rationalized by
observing that the tension varies only over distances on the order of
$\lpar$, which is at any time much larger than the correlation length
for transverse fluctuations~\cite{hallatschek-frey-kroy:05}:
\begin{equation}\label{eq:scaling-lperp}
  \lperp(t) \simeq
  \begin{cases}
    t^{1/4},&\text{ for }t\ll f^{-2}\\
    (f\,t)^{1/2},&\text{ for }t\gg f^{-2}
  \end{cases}
\end{equation}
Eq.~\eqref{eq:scaling-lperp} is easily understood as the expression of
the dynamic force balance between transverse friction forces
$r_\perp/t$ and bending $r_\perp/\lperp^4$ or tension terms $f
r_\perp/\lperp^2$, respectively, i.e., by analyzing
eq.~\eqref{eq:eom-t} on a scaling level. It displays the crossover
between ``free'' (bending-dominated) relaxation within blobs and
``forced'' (tension-driven) relaxation on larger scales.

\section{\label{sec:blob-picture}Blob picture of stretching dynamics}

\begin{figure}
  \includegraphics[width=.48\textwidth]{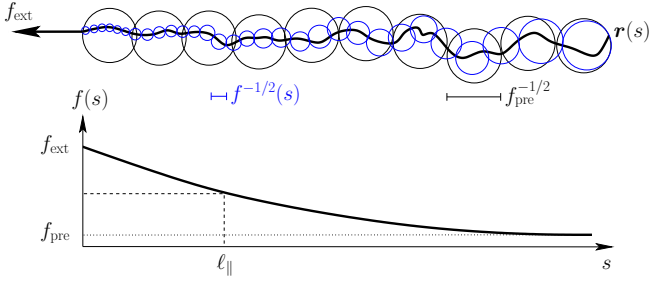}
  \caption{\label{fig:blob-picture}Pulling a prestretched filament.
    Thermal fluctuations in the contour $\bvec r(s)$ are straightened
    first in a boundary layer of width $\lpar(t)$. At any time $t$,
    blobs of size $\fpre^{-1/2}$ and $f^{-1/2}(s)$ are associated with
    the prestretching force $\fpre$ and the actual local backbone
    tension $f(s)$, respectively.  }
\end{figure}

A qualitative yet rather thorough understanding of the nonequilibrium
stretching dynamics can be gained already from the special case
$\fext\gg\fpre$, i.e., for the scenario of a sudden strong increase in
stretching force. It is useful to interpret the correlation length
$\lperp(t)$ of eq.~\eqref{eq:scaling-lperp} as an \emph{equilibration
  length} for transverse fluctuations. At any time $t$, segments of
length $\lperp$ are in equilibrium with their
surroundings~\cite{everaers-juelicher-ajdari-maggs:99,hallatschek-frey-kroy:05}.
Upon applying the external force $\fext$, the contour starts to
stretch within a boundary layer of size $\lpar$.  Decomposing the
latter into $\lpar/\lperp$ segments of length $\lperp$, the extension
$\delta$ of each segment in response to the local tension $f(s)$ can
be estimated within equilibrium theory. At this point we use the blob
picture: the prestretching force $\fpre$ and the local tension $f(s)$,
which builds up after the external force is applied, induce blobs at
different length scales. Associated with the weaker prestretching
force are large blobs of constant size $\fpre^{-1/2}$, and the
stronger local tension corresponds to small blobs of varying size
$f^{-1/2}(s)$, respectively (see fig.~\ref{fig:blob-picture}).  The
extension $\delta$ depends on ``how many'' blobs are contained in a
segment (or vice versa). Three cases can be distinguished for
$\fext\gg\fpre$:

\begin{itemize}

\item $\lperp\ll f^{-1/2}(s)$: a large number of segments $\lperp$ are
  in either type of blob; hence, bending forces dominate and the
  extension follows from linear response%
  ~\cite{mackintosh-kaes-janmey:95}: $\delta\simeq\lperp^4 f(s)/\lp$.

\item $f^{-1/2}(s)\ll\lperp\ll \fpre^{-1/2}$: segments $\lperp$ near
  the boundary are larger than the small blobs corresponding to the
  local tension.  Hence, they get almost completely stretched, and
  essentially all initially stored length is pulled out. Being still
  smaller than the large blobs of size $\fpre^{-1/2}$, these segments
  correspond to short stiff \emph{initially unstretched} chains, thus
  their extension is $\delta\simeq\lperp^2/\lp$.

\item $\fpre^{-1/2}\ll\lperp$: now the segments $\lperp$ are larger
  than any of the blobs and release the stored length of a taut
  \emph{string of blobs}.  In a segment, there are
  $\lperp/\fpre^{-1/2}$ large blobs of size $\fpre^{-1/2}$ each with
  stored length $1/(\lp\fpre)$, hence
  $\delta\simeq\lperp/(\lp\fpre^{1/2})$.
\end{itemize}

The above results for the extension $\delta$ and
eq.~\eqref{eq:scaling-lperp} for the length $\lperp$ of the segments
can now be used to estimate the total stretching
$(\lpar/\lperp)\delta$ of the chain's boundary layer. On the scaling
level, we can set $f(s)=\fext$ for the relevant segments near the
boundary. The size $\lpar$ of the boundary layer is the central
quantity still to be determined. According to its definition, it is
obtained by requiring that the total longitudinal friction on the
order of $\lpar(\lpar/\lperp)\delta/t$ equals the driving force
$\fext$, i.e., $\lpar(t)$ scales as
\begin{subequations}\label{eq:scaling-lpar}
  \begin{align}
    \label{eq:scaling-lpar-1}
    &\lp^{1/2} t^{1/8},
    &&\text{for }t\ll\fext^{-2}\\
    \label{eq:scaling-lpar-2}
    &\lp^{1/2} (\fext\, t)^{1/4},
    &&\text{for }\fext^{-2}\ll t \ll (\fext\fpre)^{-1} \\
    \label{eq:scaling-lpar-3}
    &\lp^{1/2} \fpre^{1/4} (\fext\,t)^{1/2}, &&\text{for
    }(\fext\fpre)^{-1}\ll t.
  \end{align}
\end{subequations}

The first case [eq.~\eqref{eq:scaling-lpar-1}] has already been
derived by Everaers et~al.~\cite{everaers-juelicher-ajdari-maggs:99}
(EJAM) based on a similar argument. This universal initial regime also
shows up for other force
protocols~\cite{hallatschek-frey-kroy:05,hallatschek-frey-kroy:07b},
including the case where $\fext < \fpre$. The second scaling law
[eq.~\eqref{eq:scaling-lpar-2}] has first been proposed by Seifert,
Wintz, and Nelson~\cite{seifert-wintz-nelson:96} (SWN). This
intermediate regime emerges if the large and small blobs have very
different size, i.e., either for vanishing prestretching force or for
strong force scale separation $\fext\gg\fpre$. However, \emph{only
  near the boundary} the different blobs differ in size.  Accordingly,
the contour stretches significantly only where the actual tension is
already strongly different from its bulk value $\fpre$. Thus, as
pointed out in ref.~\cite{ajdari-juelicher-maggs:97}, \emph{tension
  propagates faster than stretching is achieved}. Finally, for long
times $t\gg(\fext\fpre)^{-1}$, we get a result similar to the one by
Brochard, Buguin, and de Gennes (BBG) derived in
ref.~\cite{brochard-buguin-degennes:99}, where they assumed the
tension to be locally equilibrated. Note, however, that we obtain a
more detailed force dependence in eq.~\eqref{eq:scaling-lpar-3}.  In
sec.~\ref{sec:tension-profiles} below, this discrepancy is traced back
to a subtlety of the limit $\fpre\to 0$. Both results agree only for
$\fext\approx\fpre$.

In the latter case, i.e., if the stretching force is changed only by a
small amount $\Delta f\equiv\fext-\fpre\ll \fext$, the results for the
segment extension $\delta$ are reduced by a factor $\Delta f/\fext$.
Because the different blobs have then essentially the same size, the
force balance argument leading to eq.~\eqref{eq:scaling-lpar} (here
with the driving force $\Delta f$) yields only two cases:
$\lpar\simeq\lp^{1/2}t^{1/8}$~\cite{everaers-juelicher-ajdari-maggs:99}
for $t\ll\fext^{-2}$ and $\lpar\simeq\lp^{1/2}\fext^{3/4}t^{1/2}$%
~\cite{brochard-buguin-degennes:99} for $t\gg\fext^{-2}$. There is no
intermediate $t^{1/4}$-regime.

\begin{figure}
  \includegraphics[width=.48\textwidth]{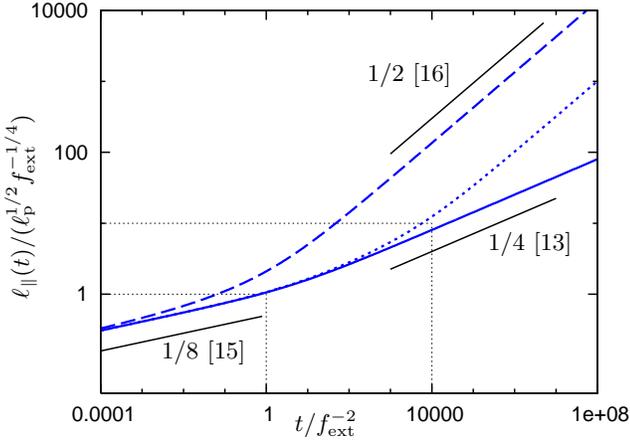}
  \caption{\label{fig:lpar-scaling} Scaling of the boundary layer size
    $\lpar(t)$ (log-log scale) for different force ratios
    $\fpre/\fext=1/2$ (long-dashed), $\fpre/\fext=10^{-4}$
    (short-dashed), and $\fpre=0$ (solid).  $\lpar$ is extracted from
    numerical solutions of eq.~\eqref{eq:pide}. The asymptotic scaling
    laws of eq.~\eqref{eq:scaling-lpar} are indicated by dark lines.
  }
\end{figure}

Summarizing the preceding discussion, we found intermediate asymptotic
scaling laws for the boundary layer size $\lpar(t)$ by estimating the
stretching of segments $\lperp$ within a blob picture and balancing
the resulting longitudinal friction with the driving force. As a rule
of thumb, the small blobs corresponding to the stronger force $f(s)$
decide whether a segment gets stretched ``just a little'' or ``much'',
and the large blobs corresponding to the weaker force $\fpre$
determine how much ``much'' actually is. To illustrate the power laws
given in eq.~\eqref{eq:scaling-lpar}, numerical results for $\lpar(t)$
are shown in fig.~\ref{fig:lpar-scaling}, where $\lpar(t)$ has been
extracted from tension profiles computed by numerically solving
eq.~\eqref{eq:pide}, see sec.~\ref{sec:numerics}. The intermediate
asymptotic scaling is clearly visible for small and large times, and
the $t^{1/4}$-regime of eq.~\eqref{eq:scaling-lpar-2} appears for very
strong force scale separation only. We proceed with a more detailed
discussion based on tension profiles to elucidate the similarities and
differences between our results and previous work.

\section{\label{sec:tension-profiles}Tension profiles}

\begin{figure}
  \includegraphics[width=.48\textwidth]{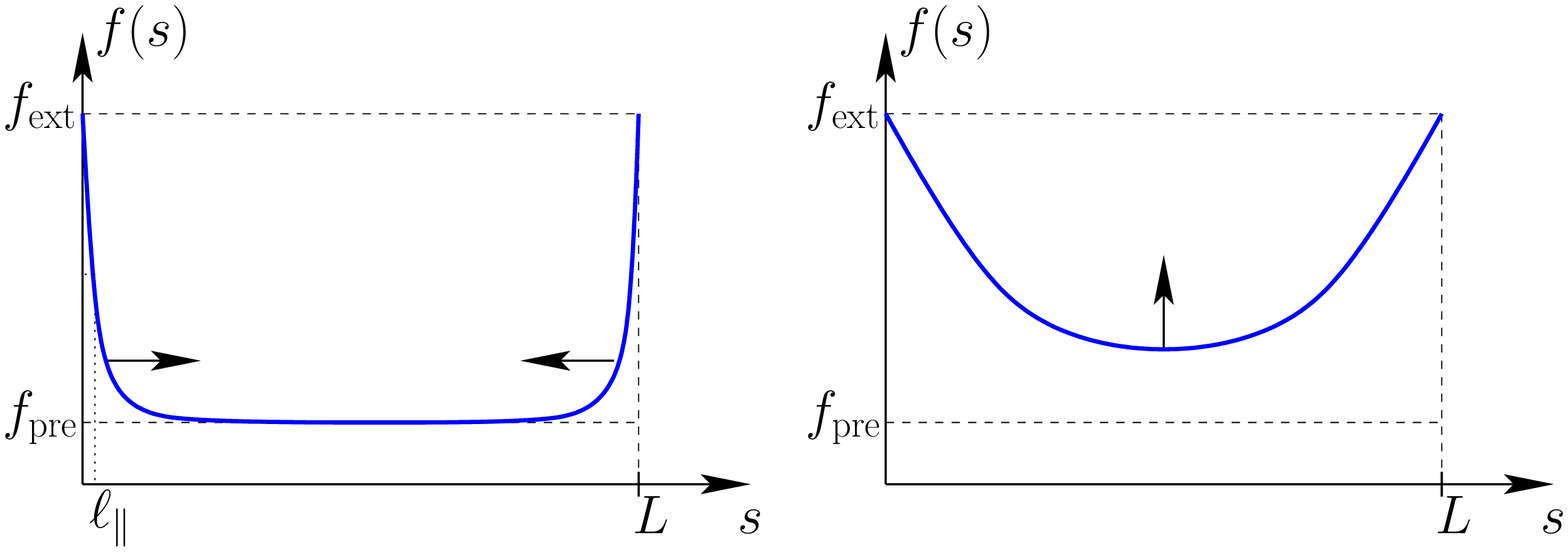}
  \caption{\label{fig:tension-profiles}Schematic tension profiles for
    $\fext > \fpre$. Left: propagation regime $t\ll\tlpar$. Near the
    ends, the tension falls off over the distance $\lpar\ll L$ and is
    constant in the bulk. Right: relaxation regime $t\gg\tlpar$. The
    tension relaxes towards its final equilibrium value
    $f\equiv\fext$. For times $t\gg t_\star$ the difference between
    $f$ and $\fext$ is small.}
\end{figure}

Instead of repeating the rigorous analysis of eq.~\eqref{eq:pide}
presented in ref.~\cite{hallatschek-frey-kroy:07b} and extending it to
deal with two force scales, we will analyze three different asymptotes
of eq.~\eqref{eq:pide} motivated by physical considerations that lead
to the quite different approximations employed
previously~\cite{everaers-juelicher-ajdari-maggs:99,seifert-wintz-nelson:96,brochard-buguin-degennes:99}.
As in sec.~\ref{sec:blob-picture}, we are primarily interested in the
case $\fext\gg\fpre$ and discuss other possibilities only
incidentally. A more systematic and complete derivation is presented
in appendix~\ref{app:analytics-asymptotes}.  Having motivated these
approximations and derived their asymptotic validity, we will find
that the partial integro-differential equation (PIDE)
[eq.~\eqref{eq:pide}] reduces to simpler differential equations that
can again be solved in two regimes (see
fig.~\ref{fig:tension-profiles}):
\begin{enumerate}
\item In the \emph{propagation regime} $t \ll \tlpar$ with $\tlpar$
  defined via $\lpar(\tlpar)=L$, the tension profile falls off over
  the (increasing) distance $\lpar(t)$. Because then $\lpar(t)\ll L$,
  we may neglect the presence of a second end, set $L\to\infty$ and
  work with a so-called ``semi-infinite'' contour.  For this regime we
  will show that, using SWN's and BBG's approaches, the resulting
  differential equations do not always lead to tension profiles with
  the required features (compare fig.~\ref{fig:tension-profiles}),
  namely that $F(s)$ decays from $\fext\,t$ (at the boundary) to
  $\fpre\,t$ (in the bulk) within a region of size $\lpar$, and that
  $F(s)\equiv\fpre\,t$ is flat in the bulk.
\item In the \emph{relaxation regime} $t\gg\tlpar$, the influence of
  the second end cannot be neglected. However, it turns out that for
  late times $t\gg t_\star$, the resulting tension profile can be
  obtained by expanding eq.~\eqref{eq:pide} about the final
  equilibrium profile: $F(s,t)=\fext\,t+\delta F(s,t)$. To lowest
  order, the correction $\delta F$ is then a simple parabola with
  constant curvature given by the right hand side of
  eq.~\eqref{eq:pide} evaluated with $F\equiv \fext\,t$. The time
  $t_\star$ is the crossover time to a regime $t\gg t_\star$ where
  $\delta F\ll\fext\,t$ is in fact small, and can therefore be defined
  via $\delta F(t_\star)\simeq \fext t_\star$. With one exception, we
  will find that $t_\star\simeq \tlpar$. Some of the relaxation
  regimes have already been discussed in
  refs.~\cite{hallatschek-frey-kroy:05,hallatschek-frey-kroy:07b} for
  related force protocols. We shift their analysis to
  appendix~\ref{app:analytics-solutions} and consider in the following
  only generically new results.
\end{enumerate}
The scaling of the crossover times $\tlpar$ and $t_\star$ between the
respective propagation and relaxation regimes will be derived along
the way and is summarized in table~\ref{tb:crossover-times} below. It
is different for the three approximations analyzed in the remainder of
this section and depends on the magnitude of $\fpre$ and $\fext$
relative to the critical force $\fc=\lp^2/L^4$.  Similar to the
well-known Euler buckling force $f^*\propto L^{-2}$ that represents a
critical threshold above which forces significantly disturb the
\emph{transverse} dynamics~\cite{granek:97,hallatschek-frey-kroy:05},
the force $\fc$ can be seen as threshold force for the
\emph{longitudinal} dynamics.

\subsection{\label{sec:linear-regime}Linear approximation}
The two limits discussed above arise quite naturally within the
\emph{linear approximation} to eq.~\eqref{eq:pide}, which is similar
to the approach by EJAM~\cite{everaers-juelicher-ajdari-maggs:99}. It
consists in treating the prestretching force $\fpre$ as well as the
tension $F$ as small perturbations with respect to the dynamically
relevant bending contributions.  From eq.~\eqref{eq:scaling-lperp} and
its interpretation we infer that this approximation is justified in
the limit $t\ll f^{-2}$. Because the tension $f$ varies between
$\fext$ at the boundary and $\fpre$ in the bulk, this certainly holds
if $t\ll\min(\fext^{-2},\fpre^{-2})$. In this case, we approximate the
response function [eq.~\eqref{eq:chi-perp}] by
$\chi_\perp(q;t,t')\approx \e^{-2 q^4 t}(1-2 q^2 F)$ and the initial
mode spectrum [eq.~\eqref{eq:initial-spectrum}] by
$(q^2+\fpre)^{-1}\approx q^{-2}(1-\fpre/q^2)$, and obtain
eq.~\eqref{eq:pide} to linear order in $F$ and $\fpre$ as:
\begin{multline}\label{eq:pide-linearized}
  \pd_s^2 F(s,t) \approx \int_0^\infty\!\frac{\td q}{\pi\lp}\Bigg[
  -\frac{\fpre}{q^4}\left(1-\e^{-2 q^4 t}\right) \\
  + 2 F(s,t) - 4 q^4 \int_0^t\!\td t'\,F(s,t')\e^{-2 q^4
    (t-t')}\Bigg].
\end{multline}
Using the Laplace transform $\tilde F(s,z)=\mathcal{L}\{F(s,t)\}$,
this reads
\begin{equation}
  \pd_s^2\tilde F(s,z)=\int_0^\infty\!\frac{\td q}{\pi\lp}\left[
    -\frac{2 \fpre}{z(z+2q^4)} + \tilde F(s,z) \frac{2 z}{z+2 q^4}\right],
\end{equation}
which, after performing the $q$-integral, reduces to:
\begin{equation}\label{eq:pide-linearized-laplace}
  \lambda^2\,\pd_s^2\tilde F = \tilde F-\frac{\fpre}{z^2}.
\end{equation}
Here, $\lambda(z)=2^{3/8}\lp^{1/2} z^{-1/8}$ is a dynamic length scale
denoting the size of spatial variations in $\tilde F(s,z)$. It is
therefore directly related to $\lpar(t)$. With the boundary conditions
$\tilde F(0,z)=\tilde F(L,z)=\fext/z^2$, the solution to
eq.~\eqref{eq:pide-linearized-laplace} reads
\begin{equation}\label{eq:laplace-solution}
  \tilde F(s,z)= \frac{\Delta f}{z^2} \frac{\cosh[(L-2 s)/2\lambda]}
  {\cosh(L/2\lambda)}+\frac{\fpre}{z^2},
\end{equation}
where $\Delta f\equiv\fext-\fpre$ is the force difference.

If $L\gg\lambda$, the tension profile eq.~\eqref{eq:laplace-solution}
varies only close to the boundaries, as it is characteristic for the
propagation regime. Near $s=0$, it simplifies to
\begin{equation}\label{eq:laplace-solution-propagation}
  \tilde F(s,z) \approx\frac{\Delta f}{z^2} \e^{-s/\lambda} + \frac{\fpre}{z^2},
\end{equation}
which can be backtransformed~\cite{hallatschek-frey-kroy:07b} to
\begin{equation}\label{eq:solution-linear-propagation}
  F(s,t) = \Delta f\,t\, \phi(s/\lpar(t)) + \fpre t,
\end{equation}
where $\phi(\xi)\approx \exp[-2^{-3/8}\xi/\EGamma{15/8}]$ is a scaling
function that depends only on the ratio $\xi=s/\lpar(t)$ with a
boundary layer size $\lpar(t)=\lp^{1/2}t^{1/8}$ that scales as
predicted in eq.~\eqref{eq:scaling-lpar-1}. The requirement
$L\gg\lambda$ translates into $t\ll\tlpar$ with $\tlpar\simeq
L^8/\lp^4$.

The relaxation regime corresponds to the complementary limit
$L\ll\lambda$, or equivalently $t\gg\tlpar$. Here we find, as
anticipated, that the tension $F(s,t)$ differs for $t\gg t_\star$ from
the flat profile $\fext\,t$ only by a subdominant term $\delta F\simeq
\fext t^{3/4} L^2/\lp$ with a (trivial) parabolic $s$-dependence.  By
evaluating $\delta F(t_\star)=\fext t_\star$ we find $t_\star\simeq
L^8/\lp^4\simeq \tlpar$. See appendix~\ref{app:analytics-solutions}
for details.

\subsection{\label{sec:taut-string}Taut-string approximation}
In this approximation introduced by
SWN~\cite{seifert-wintz-nelson:96}, bending and thermal forces are
neglected after the preparation of an initial equilibrium
configuration. As discussed below eq.~\eqref{eq:scaling-lperp},
bending contributions are locally negligible against tension in the
long-time limit $t\gg f^{-2}$. The transverse displacement modes
effectively obey $\pd_t\rperp= -q^2 f \rperp$, see
eq.~\eqref{eq:eom-t-modes}. The stored length density relaxes as
$\varrho=\sum_q \varrho_0(q)\exp[-2 q^2 F]$, and eq.~\eqref{eq:eom-l}
reads~\cite{hallatschek-frey-kroy:07b}:
\begin{equation}
  \pd_{s}^2 F = \int_0^\infty\!\!\frac{\td q}{\pi\lp} 
  \frac{1-\e^{-2 q^2 F}}{q^2 + \fpre} .
\end{equation}
Depending on the magnitude of the product $\fpre F$ (which is
dimensionless in our units), this has two asymptotes:
\begin{subequations}\label{eq:ode-swn}
  \begin{empheq}[left={\displaystyle{\pd_s^2 F \sim
        \empheqlbrace}}]{align}
    \label{eq:ode-swn-1}
    &\sqrt{2 F/\pi\lp^2},&\text{if }\fpre F\ll 1,\\
    \label{eq:ode-swn-2}
    &1/(2\fpre^{1/2}\lp),&\text{if }\fpre F\gg 1.
  \end{empheq}
\end{subequations}
These will be discussed only for the propagation regime.

Estimating $ F\simeq\fext t$ near the boundary, the first asymptotics
is realized for intermediate times $\fext^{-2}\ll
t\ll(\fpre\fext)^{-1}$.  The differential
equation~\eqref{eq:ode-swn-1} is
solved~\cite{hallatschek-frey-kroy:07b} by the scaling ansatz
$F(s,t)\equiv\fext t\,\phi(s/\lpar(t))$, with the boundary layer size
$\lpar(t)=\lp^{1/2}(\fext\,t)^{1/4}$ as in
eq.~\eqref{eq:scaling-lpar-2}, which implies $\tlpar\simeq
L^4/(\lp^2\fext)$.  This scaling ansatz gives a tension that decays
within a length $\lpar(t)$ as expected, but does it meet our
expectations in the bulk?  If $\fpre=0$, corresponding to the
``pulling''-scenario of ref.~\cite{hallatschek-frey-kroy:05}, the
scaling form $F=\fext t\,\phi(\xi)$ with $\xi=s/\lpar(t)$ can smoothly
be extended to the bulk tension, which obeys $F\equiv \pd_s^2
F\equiv0$~\cite{hallatschek-frey-kroy:07b}:
\begin{equation}\label{eq:solution-swn-propagation}
  \phi(\xi)=\left[1-\frac{\xi}{(72\pi)^{1/4}}\right]^4 
  \Theta\left((72\pi)^{1/4}-\xi\right).
\end{equation}
If $\fpre$ is finite, however, the bulk tension has nonzero magnitude
and eq.~\eqref{eq:ode-swn} would yield nonzero curvature for the
expectedly flat bulk profile. Hence, for finite $\fpre$ the
taut-string approximation is not valid along the whole contour, but
only in a finite region near the boundary. \emph{In this case it does
  not represent an intermediate asymptotic regime of
  eq.~\eqref{eq:pide}.}

The second asymptotics~\eqref{eq:ode-swn-2}, realized for late times
$t\gg(\fpre\fext)^{-1}$, implies zero curvature for the tension
$f=\pd_t F$ everywhere. The relaxation changes its character: most of
the initially excited modes have relaxed (the long-wavelength
contributions have been ``cut off'' by the prestretching force from
the beginning). This corresponds to an almost completely stretched
contour under linearly decreasing tension in the boundary layer.
However, eq.~\eqref{eq:ode-swn-2} is not sufficient to describe the
specific shape of the associated tension profiles which have to be
constant in the bulk. One needs to include thermal
noise.\footnote{SWN's ``noiseless''
  simulations~\cite{seifert-wintz-nelson:96} for a prestretched chain
  yield the correct scaling $\lpar(t)\propto t^{1/2}$ and linear
  tension profiles, but (as their scaling argument indicates) a
  straight string drawn through a viscous solvent gives a linear
  profile, irrespective of thermal noise.}

\subsection{\label{sec:quasi-static}Quasi-static approximation}
The preceding paragraph revealed a problem with the taut-string
approximation for $f^2 t\gg1$ and $\fpre F\gg 1$: the right hand side
of eq.~\eqref{eq:ode-swn-2} becomes time-independent and the
deterministic relaxation saturates, so that the subsequent dynamics is
purely of stochastic origin. This allows to simplify it as a
quasi-equilibrium process. Following the approach of
BBG~\cite{brochard-buguin-degennes:99}, one can assume the filament to
be equilibrated under the local tension $f(s,t)=\pd_t F(s,t)$. From
the continuum approximation to eq.~\eqref{eq:initial-spectrum}, the
stored length density is $\varrho=\int\frac{\td q}{\pi\lp}(q^2+\pd_t
F)^{-1}=(\pd_t F)^{-1/2}/(2\lp)$. According to
eq.~\eqref{eq:eom-l-integrated}, this adds a small but relevant
contribution to
eq.~\eqref{eq:ode-swn-2}~\cite{hallatschek-frey-kroy:07b}:
\begin{equation}\label{eq:ode-bbg}
  \pd_{ s}^2  F = \frac{1}{2\lp}\left[\fpre^{-1/2}
    -(\pd_t F)^{-1/2}\right].
\end{equation}
By taking a time derivative we get
\begin{equation}\label{eq:ode-bbg-derivative}
  \pd_{s}^2 f=\frac{\pd_t f}{4\lp f^{3/2}}.
\end{equation}
We discuss the quasi-static approximation both for the propagation and
the relaxation regime.

\begin{figure}
  \includegraphics[width=.48\textwidth]{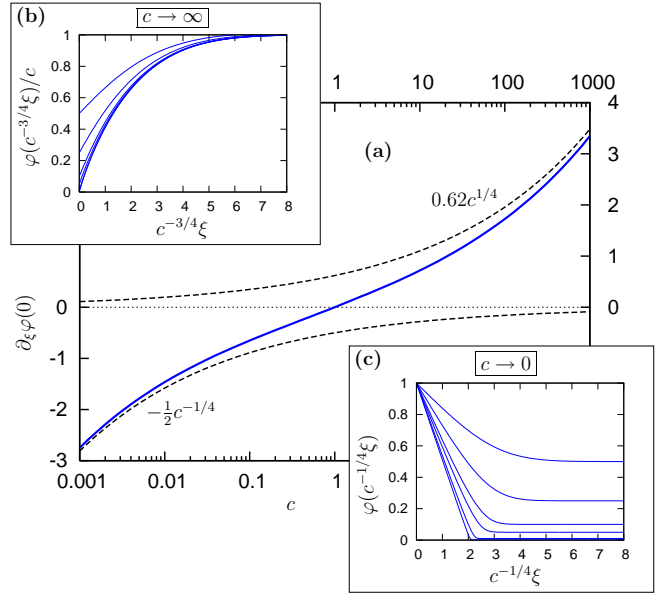}
  \caption{\label{fig:ode-bbg}Numerical solutions of
    eq.~\eqref{eq:ode-bbg-scaling} for the scaling function
    $\varphi(\xi)$ of the tension profile $f(s,t)$.  (a) Plot of the
    slope $\pd_\xi\varphi(0)$ at the boundary (solid line) vs.\ force
    ratio $c=\fpre/\fext$ (log-scale) with analytical asymptotes for
    $c\to0$ and $c\to\infty$, respectively (dashed lines). (b) The
    scaling functions ($c=2,4,10,20,100,1000$ from top to bottom)
    asymptotically collapse for $c\to \infty$ if the abscissa is
    rescaled by $c^{-3/4}$ and the ordinate by $c^{-1}$. (c) The
    scaling functions ($c=.5,.25,.1,.05,.01,.001$ from top to bottom)
    asymptotically collapse onto a piecewise linear profile for $c\to
    0$ if the abscissa is rescaled by $c^{-1/4}$.}
\end{figure}

\paragraph{Propagation regime.}
Here we insert the scaling ansatz $f(s,t)\equiv\fext \varphi(\xi)$
with $\xi=s/\lpar^*(t)$ and the tentative scaling
$\lpar^*(t)=\lp^{1/2}\fext^{3/4}t^{1/2}$ as proposed by BBG. This
leads to an ordinary differential equation
\begin{equation}\label{eq:ode-bbg-scaling}
  \pd_\xi^2\varphi = -\frac{1}{8}\xi\varphi^{-3/2}\pd_\xi\varphi.
\end{equation}
Boundary conditions are $\varphi(0)=1$ and $\varphi(\xi\to\infty)=c$
where $c\equiv\fpre/\fext$ is the force ratio. This is the same
equation as eq.~(18) of ref.~\cite{brochard-buguin-degennes:99}, safe
for a factor $\frac{1}{4}$ due to a slightly different definition of
the scaling variable $\xi$.  We have solved it numerically by a
shooting method.  Starting with $\varphi(0)=1$, it is integrated
forward, adjusting the slope $\pd_\xi\varphi(0)$ at the left boundary
in order to fulfill the second boundary condition. The dependence of
the slope $\pd_\xi\varphi(0)$ on the force ratio $c$ is shown in
fig.~\ref{fig:ode-bbg}(a), together with the asymptotes for
$c\to\infty$ and $c\to 0$, respectively. These shall now be analyzed
in more detail.

The case $c > 1$ corresponds to a stretching force $\fext < \fpre$
that is decreased at $t=0$, i.e., a ``sudden release''-scenario as in
refs.~\cite{brochard-buguin-degennes:99,hallatschek-frey-kroy:05,hallatschek-frey-kroy:07b}.
Since $\fpre$ is the relevant force scale, the scaling ansatz is more
appropriately written as $f(s,t)=c\,\fext\varphi(c^{-3/4}\xi)$. Now we
can safely take the limit $c\to\infty$ ($\fext\to 0$) describing the
case where the force is switched off completely,\footnote{If for
  strongly prestretched ``flexible WLC'' with $\lp\ll L$ and
  $\fpre\gg\lp^{-2}$ the external force is decreased too much
  ($\fext\lesssim\lp^{-2}$), the weakly-bending approximation
  eventually breaks down near the ends as the tension decreases. This
  restricts the applicability of our analysis to the dynamics of those
  observables that are insensitive to boundary
  contributions~\cite{hallatschek-frey-kroy:07b}.}  because the
artificial $c^{1/4}$-divergence in $\pd_\xi\varphi(0)$ has been
removed.  The resulting boundary layer size scales like
$\lpar(t)\simeq \lp^{1/2}\fpre^{3/4} t^{1/2}$ as in
ref.~\cite{brochard-buguin-degennes:99} and the corresponding scaling
functions, shown in fig.~\ref{fig:ode-bbg}(b), smoothly converge to
the solutions depicted in
refs.~\cite{brochard-buguin-degennes:99,hallatschek-frey-kroy:07b}.

The opposite limit $c\to 0$ is more subtle. Although it is not obvious
from eq.~\eqref{eq:ode-bbg-scaling}, we may not simply set $\fpre=0$
as in ref.~\cite{brochard-buguin-degennes:99}, since then the
condition $\fpre F\gg 1$ in eq.~\eqref{eq:ode-swn-2} cannot be met.
Moreover, considering the limit $c\to 0$ with $\fpre>0$ fixed instead,
our numerical results (fig.~\ref{fig:ode-bbg}(a)) indicate that the
initial slope $\pd_\xi\varphi(0)$ diverges like $c^{-1/4}$. This
suggests the scaling variable $\eta\equiv c^{-1/4}\xi=s/\lpar(t)$ with
the new boundary layer scaling
$\lpar(t)=c^{1/4}\lpar^*(t)=\lp^{1/2}\fpre^{1/4}(\fext t)^{1/2}$ as
anticipated in eq.~\eqref{eq:scaling-lpar-3}.  We can further
rationalize this asymptotic scaling by inserting the improved scaling
ansatz $F(s,t)=\fext t\,\phi(\eta)$ into eq.~\eqref{eq:ode-bbg}, which
gives $\pd_\eta^2\phi(\eta)=\frac{1}{2}+\mathcal{O}(c^{1/2})$. In the
limit $c\to 0$, the scaling function $\phi$ is parabolic in the
boundary region $\eta\in[0,2]$, and the scaling function
$\varphi=\phi-\frac{1}{2}\eta\pd_\eta\phi=1-\frac{1}{2}\eta$ for the
tension $f$ becomes linear. Numerical solutions to
eq.~\eqref{eq:ode-bbg-scaling} in terms of $\eta=\xi c^{-1/4}$ in fact
tend towards a piecewise linear function with a kink fixed at $\eta=2$
(see fig.~\ref{fig:ode-bbg}(c)), hence the kink is moving towards
vanishing $\xi=c^{1/4}\eta$ as $c\to 0$. This reveals that the limit
$c\to 0$ cannot properly be taken in terms of the original scaling
variable $\xi$, because eq.~\eqref{eq:ode-bbg} and therefore
eq.~\eqref{eq:ode-bbg-scaling} become invalid if $c=\fpre/\fext=0$.

\paragraph{Relaxation regime.}
In the relaxation regime $t\gg t_\star$ we expect the tension to
deviate only slighly from the flat equilibrium profile
$F\equiv\fext\,t$, see fig.~\ref{fig:tension-profiles}. We therefore
write $F(s,t)=\fext\,t+\delta F(s,t)$ and expand
eq.~\eqref{eq:ode-bbg}, which gives the \emph{time-independent}
correction:
\begin{equation}\label{eq:deltaF-premature-saturation}
  \delta F = \frac{1}{4\lp}\left[\fpre^{-1/2}-\fext^{-1/2}\right] s(s-L).
\end{equation}
This result has two significant implications. First of all, in order
to estimate the time $t_\star$ beyond which $\delta F$ becomes small
compared to $\fext\,t$, we evaluate $\delta F(t_\star)=\fext\,t_\star$
and find that not always $t_\star\simeq\tlpar$, see the last two rows
of table~\ref{tb:crossover-times}. We find for $\fpre\gg\fext$ and
$\fpre\gg \fc$ another intermediate regime $\tlpar\ll t\ll t_\star$ of
\emph{homogeneous tension relaxation}, where the tension $f$ has a
more complicated than just parabolic spatial dependence and the simple
ansatz $F(s,t)=\fext\,t+\delta F(s,t)$ for the relaxation regimes does
not work. This has been attributed~\cite{hallatschek-frey-kroy:05} to
the fact that longitudinal friction may play an essential role also
for times $t\gg\tlpar$ beyond the propagation regime and significantly
slow down the relaxation. Safe for small correction terms (see
appendix~\ref{app:analytics-solutions}), our results for the
homogeneous relaxation regime are identical to the ``sudden
release''-case $\fext=0$ discussed in
ref.~\cite{hallatschek-frey-kroy:07b}.

Secondly, a time-independent correction $\delta F$ of the integrated
tension yields an actual tension $f=\pd_t F$ that (to leading order)
does not deviate at all from the flat equilibrium profile
$f\equiv\fext$.  The relaxation dynamics saturates \emph{prematurely}
to equilibrium, i.e., at times $t\simeq t_\star$ long before the
longest bending mode has relaxed (see also eq.~\eqref{eq:tlperp}
below).  This behavior arises within a scenario where a strong
stretching force $\fpre\gg\fc$ is changed to $\fext\gg\fc$ but remains
strong.  Hence, there is no need to destroy (create) long-wavelength
contributions in the mode spectrum of the stored length (see
eq.~\eqref{eq:initial-spectrum}) as in the
``pulling''(``release'')-scenario of
ref.~\cite{hallatschek-frey-kroy:05}.  By taking the next-to-leading
order term in the quasi-static approximation [eq.~\eqref{eq:ode-bbg}]
(cf.\ appendix~\ref{app:analytics-solutions}), we find an improved
correction term that captures the slow diffusive relaxation of the
remaining long-wavelength modes:
\begin{multline}\label{eq:deltaF-premature-saturation-improved}
  \delta F(s,t)=\frac{1}{4\lp}\left(\fpre^{-1/2}-\fext^{-1/2}\right)s(s-L) \\
  - \frac{1}{4\lp}\frac{\fpre^{-1}-\fext^{-1}}
  {\sqrt{2\pi}(\fext\,t)^{1/2}}s(s-L).
  \tag{\ref{eq:deltaF-premature-saturation}'}
\end{multline}

\begin{table}
  \begin{center}
    \renewcommand{\arraystretch}{1.5}
    \begin{tabular}{c||c|c}
      \hline
      & $\tlpar$ & $t_\star$ \\
      \hline\hline
      $\fpre\ll\fext\ll\fc$ & \multicolumn{2}{c}{\multirow{2}{*}{$L^8/\lp^4$}} \\
      \cline{1-1}
      $\fext\ll\fpre\ll \fc$ & \multicolumn{2}{c}{} \\
      \hline
      $\fpre\ll\fc\ll\fext$ & \multicolumn{2}{c}{$L^4/(\lp^2\fext)$} \\
      \hline
      $\fc\ll\fpre\ll\fext$ & \multicolumn{2}{c}{$L^2/(\lp\fpre^{1/2}\fext)$} \\
      \hline
      $\fext\ll\fc\ll\fpre$ & \multirow{2}{*}{$L^2/(\lp\fpre^{3/2})$} & $L^8/\lp^4$ \\
      \cline{1-1}\cline{3-3}
      $\fc\ll\fext\ll\fpre$ & & $L^2/(\lp\fext^{3/2})$ \\
      \hline
    \end{tabular}
  \end{center}
  \caption{\label{tb:crossover-times}
    Crossover times $\tlpar$ and $t_\star$ depending on the magnitude of 
    $\fpre$ and $\fext$ relative to the critical force $\fc=\lp^2/L^4$.
  }
\end{table}

Having discussed some of the asymptotic solutions to
eq.~\eqref{eq:pide} resulting from the linear, the taut-string, and
the quasi-static approximation, respectively, we now present a
numerical approach to solve eq.~\eqref{eq:pide}.

\section{\label{sec:numerics}Numerical approach}

While the intermediate asymptotes of eq.~\eqref{eq:pide} serve to
expose the relevant physics involved in the relaxation process,
experiments on the biopolymers we have in mind are usually performed
in intermediate, non-asymp\-to\-tic parameter ranges.  Therefore we
devised a numerical scheme to solve the nonlinear PIDE
[eq.~\eqref{eq:pide}] under very general initial and boundary
conditions, allowing eq.~\eqref{eq:pide} to be applied to far more
scenarios than just the ones analyzed in this work.

Numerical solutions of eq.~\eqref{eq:pide} are written in terms of
dimensionless variables. Here, we use a characteristic length scale
$\sc$ specified when necessary, and the time and force scales
$\tf=\fext^{-2}$ and $\fext$, respectively. The integrated tension is
written as
\begin{equation}\label{eq:numerical-scaling-ansatz-F}
  F(s,t) \equiv \fext\, t\, \Phi\left(\frac{s}{\sc},\frac{t}{\tf}\right),
\end{equation}
and the actual tension profile is then extracted via
\begin{equation}\label{eq:numerical-scaling-ansatz-f}
  f(s,t)=\fext\,\varphi(\sigma,\tau)\equiv\fext\,\pd_\tau[\tau\Phi(\sigma,\tau)].
\end{equation}
Inserting the scaling form [eq.~\eqref{eq:numerical-scaling-ansatz-F}]
into eq.~\eqref{eq:pide} and rescaling $q=\tilde q\sqrt{\fext}$ and
$t'=z t$, we are left with
\begin{multline}\label{eq:pide-numerical}
  \Lambda^2\, \pd_\sigma^2 \Phi(\sigma,\tau) =
  \int_0^\infty\!\frac{\td\tilde q} {\pi}\Bigg[ \frac{1-\e^{-2\tilde
      q^2\tau[\tilde q^2+\Phi(\sigma,\tau)]}}
  {\tau(\tilde q^2+c)} \\
  - 2\tilde q^2 \int_0^1\!\td z\,\e^{-2 \tilde q^2 \tau[\tilde q^2
    (1-z) + \Phi(\sigma,\tau)-z\Phi(\sigma,\tau z)]}\Bigg].
\end{multline}
The remaining parameters are
\begin{equation}\label{eq:numerical-scaling-ansatz-parameters}
  \Lambda^2=\lp/(\fext^{1/2}\sc^2)\text{ and } c=\fpre/\fext.
\end{equation}
For the ``semi-infinite'' polymers with $L\to\infty$ we use
$\Lambda^2=1$ by choosing $\sc=\lp^{1/2}\fext^{-1/4}$, while in the
finite case the appropriate choice is $\sc=L$.

Our strategy is as follows. Given the initial condition
$\varphi(\sigma,0)\equiv\varphi_0(\sigma)$ and appropriate linear
boundary conditions for the respective scenario, we introduce a
discretized time coordinate $\tau_n$ for $n=0,\ldots N$. At each time
step $\tau_n$ we obtain a two-point boundary value problem:
\begin{subequations}\label{eq:bvp}
  \begin{equation}
    \label{eq:bvp-bulk}
    \pd_\sigma^2 \Phi(\sigma,\tau_n) = G [\Phi(\sigma,\tau_k)_{k\leq n}].
  \end{equation}
  The nonlinear term $G$ depends also on ``earlier'' solutions
  $\Phi(\sigma,\tau_k)$ with $k\leq n$. We allow arbitrary linear
  boundary conditions incorporated via 6 coefficients $\alpha_{jk}$:
  \begin{align}
    \label{eq:bvp-bc-left}
    \alpha_{00} \Phi(0,\tau_n) + \alpha_{01}\pd_\sigma\Phi(0,\tau_n)
    &= \alpha_{02} \\
    \label{eq:bvp-bc-right}
    \alpha_{10} \Phi(\sigma_{M},\tau_n) +
    \alpha_{11}\pd_\sigma\Phi(\sigma_{M},\tau_n) &=\alpha_{12}.
  \end{align}
\end{subequations}
These boundary value problems [eq.~\eqref{eq:bvp}] are solved by
transforming to a system of nonlinear equations using the discretized
arc\-length coordinate $\sigma_m$ for $m=0,\ldots M$, and discrete
representations for the differential operators $\pd_\sigma$ and
$\pd_\sigma^2$. For details see Appendix~\ref{app:numerics}.

Given numerical solutions to eq.~\eqref{eq:pide}, it is
straightforward to obtain results for pertinent observables.

\section{\label{sec:deltar}Results for the change in projected length}

The observable of highest interest in experiments is certainly the
change in end-to-end distance. Since the sudden change in stretching
force (from $\fpre$ to $\fext$) induces the creation or destruction of
stored length, and since the total contour length is conserved, we can
simply evaluate the change in end-to-end distance (approximated by its
projection onto the longitudinal reference axis) by integrating up the
total difference in stored length:
\begin{equation}\label{eq:deltar}
  \begin{split}
    \Delta\overline R_\parallel(t) &= -\int_0^L\!\td s\,
    [\varrho(s,t)-\varrho(s,0)] \\
    &= \int_0^L\!\td s\, \pd_s^2 F(s,t) \\
    &= F'(L,t)-F'(0,t).
  \end{split}
\end{equation}
Here we have used eq.~\eqref{eq:eom-l-integrated} to relate the change
in stored length to the curvature of the integrated tension $F$. The
notation $\Delta \overline R_\parallel(t)$ serves as a remainder of
the fact that $\Delta \overline R_\parallel(t)$ includes only bulk
contributions. The ``real'' change in projected end-to-end distance
$\ave{\Delta R_\parallel(t)}$ measured in experiments also contains
end contributions that depend on the choice of boundary conditions
which we have neglected in our approximate mode decomposition of
eq.~\eqref{eq:eom-t}. However, a careful
analysis~\cite{hallatschek-frey-kroy:07b} shows that for hinged or
clamped ends and a sufficiently large prestretching force $\fpre\gg
L^{-2}$ these contributions vanish. For the ``semi-infinite'' polymers
discussed above we neglected the presence of a second end; here we can
exploit the symmetry of the setup and use $\Delta\overline
R_\parallel(t) = -2 F'(0,t)$.

\subsection{\label{sec:analytical-results}Analytical results}
Given the two forces $\fext$ and $\fpre$, and the additional force
scale $\fc$, the sequence of asymptotic regimes realized for a
specific choice of parameters depends on their respective ratios and
is in general quite complicated, see also
table~\ref{tb:crossover-times}. Relaxing the condition $\fext > \fpre$
of sec.~\ref{sec:blob-picture}, we chose to first illustrate
schematically the general dependence on the force ratio $\fpre/\fext$
since the systematic investigation of its influence is the major
novelty of this work. With our numerical results, we will then
demonstrate the dependence on the polymer length (more specifically,
on the parameter $\Lambda^2$ of
eq.~\eqref{eq:numerical-scaling-ansatz-parameters}) for fixed force
ratio.

\begin{figure}
  \includegraphics[width=.48\textwidth]{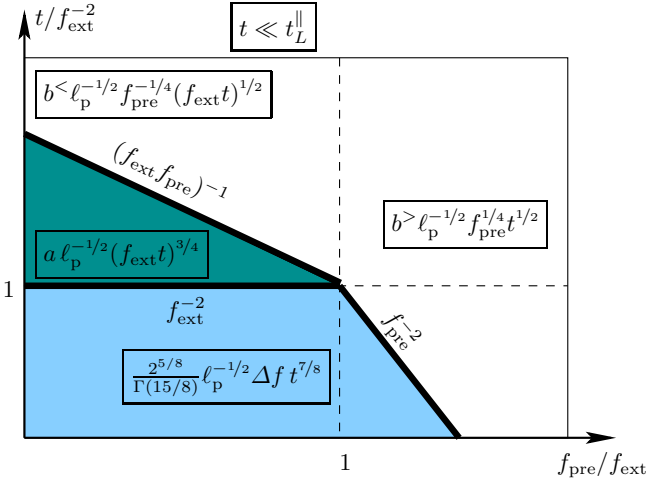}
  \caption{\label{fig:phase-diagram-propagation}Regimes of
    intermediate asymptotics (separated by thick black lines) for the
    bulk contribution to the change in end-to-end distance
    $\Delta\overline R_\parallel(t)$ (boxed formulas) in the
    propagation regime $t\ll\tlpar$; time $t/\fext^{-2}$ vs.\ force
    ratio $\fpre/\fext$ (log-log scale). Asymptotic values for the
    prefactors are $b^{<}\sim 2$, $b^{>}\sim -2.48$, and $a=4
    (2/\pi)^{\nicefrac{1}{4}}/\sqrt{3}$. See text for explanation.}
\end{figure}

\begin{figure}
  \includegraphics[width=.48\textwidth]{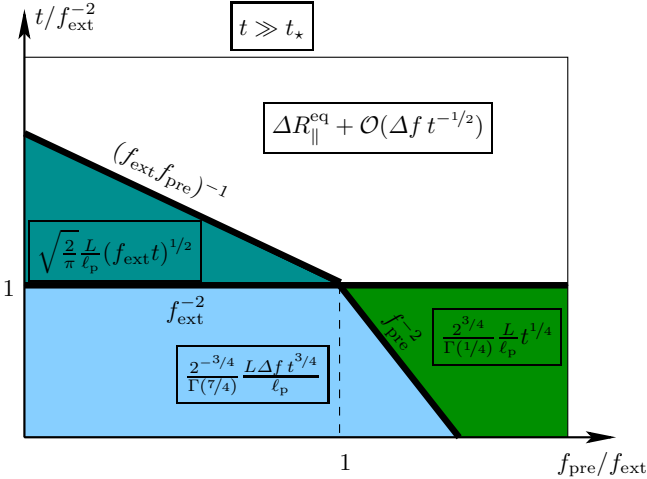}
  \caption{\label{fig:phase-diagram-relaxation}Regimes of intermediate
    asymptotics for $\Delta \overline R_\parallel(t)$ as in
    fig.~\ref{fig:phase-diagram-propagation}, but here for the
    relaxation regime $t\gg t_\star$. See text for explanation.}
\end{figure}

Using eq.~\eqref{eq:deltar}, we summarized the asymptotic growth laws
for $\Delta \overline R_\parallel(t)$ resulting from the tension
profiles computed in sec.~\ref{sec:tension-profiles} in two phase
diagrams.  We illustrate the growth laws for $\Delta \overline
R_\parallel(t)$ as boxed formulas in their respective asymptotic
regime depending on the force ratio $\fpre/\fext$ in
fig.~\ref{fig:phase-diagram-propagation} for the propagation regime
($t\ll\tlpar$) and in fig.~\ref{fig:phase-diagram-relaxation} for the
relaxation regime ($t\gg t_\star$), respectively. Given a specific
force ratio, the evolution of $\Delta \overline R_\parallel(t)$
corresponds to a vertical line through
fig.~\ref{fig:phase-diagram-propagation} until $t\simeq\tlpar$ where
it crosses over to the relaxation regimes ($t\gg t_\star$) depicted in
fig.~\ref{fig:phase-diagram-relaxation}, since in most cases
$\tlpar\simeq t_\star$.

Let us first turn to the propagation regimes of
fig.~\ref{fig:phase-diagram-propagation}. In the universal initial
regime (light shaded) the scaling is linear in the force difference
$\Delta f=\fext-\fpre$, see
eq.~\eqref{eq:solution-linear-propagation}; it is followed by a
quasi-static regime (white) with different force scaling for
asymptotically small $(<)$ and large $(>)$ force ratio; in these
limits, we find analytical values for the prefactors $b^>$ and $b^<$
as shown in the figure caption. The general formula follows from
eq.~\eqref{eq:ode-bbg} and reads
\begin{equation}
  \Delta \overline R_\parallel(t) = -4\pd_\xi\varphi(0)
  \lp^{-1/2}\fext^{1/4}t^{1/2},
\end{equation}
where the dependence of the slope $\pd_\xi\varphi(0)$ on the force
ratio $c=\fpre/\fext$ is depicted in fig.~\ref{fig:ode-bbg}. The
intermediate taut-string region (dark shaded) emerges only for very
small force ratio. Here the analytical value for the prefactor $a$
applies to the case $\fpre=0$.

Except for the case $\fpre\gg\fext,\fc$, the propagation-relaxation
crossover at $t\simeq \tlpar$ leads directly to the regimes depicted
in fig.~\ref{fig:phase-diagram-relaxation}. Here we find again a
short-time regime linear in $\Delta f$, followed by two different
intermediate regimes: the one with small $\fpre/\fext$ connects to the
``pulling''-case ($\fpre=0$), the other with large force ratio
connects to the ``release''-case ($\fext=0$) of
ref.~\cite{hallatschek-frey-kroy:07b}, respectively. In the final
regime $t\gg t_\star$ with $\fext,\fpre\gg\fc$, the to leading order
time-independent correction $\delta F$ to the equilibrium tension
profile [eq.~\eqref{eq:deltaF-premature-saturation}] implies via
eq.~\eqref{eq:deltar} that $\Delta \overline R_\parallel(t)$
prematurely attains its equilibrium value
\begin{equation}
  \Delta R_\parallel^\text{eq}=\frac{L}{2\lp} \left[\fpre^{-1/2}-\fext^{-1/2}\right].
\end{equation}
From eq.~\eqref{eq:deltaF-premature-saturation-improved} we find the
subdominant correction term
\begin{equation}\label{eq:deltar-premature-saturation}
  \Delta \overline R_\parallel(t) =\Delta R_\parallel^\text{eq}-
  \frac{\Delta f\,L}{\sqrt{32\pi} \lp \fpre \fext^{3/2} t^{1/2}}.
\end{equation}

The intermediate regime of homogeneous tension relaxation $\tlpar\ll
t\ll t_\star$ is discussed in ref.~\cite{hallatschek-frey-kroy:07b}
and Appendix~\ref{app:analytics-solutions}.  For $\fpre\gg \fc,\fext$
we obtain from eqs.~(\ref{eq:h-characteristics-slope},\ref{eq:deltar})
the expansion:
\begin{equation}\label{eq:deltar-hr}
  \Delta\overline R_\parallel(t) = -\left(\frac{18 L t}{\lp^2}\right)^{1/3}
  + L\, \mathcal{O}\left((t/t_\star)^{2/3}\right).
\end{equation}

\subsection{\label{sec:numerical-results}Numerical results}

\begin{figure}
  \psfrag{XLABEL}{$t/\fext^{-2}$} \psfrag{YLABEL}{$L\fext\Delta
    \overline R_\parallel(t)$} \centerline{\rotatebox{270}{
      \includegraphics{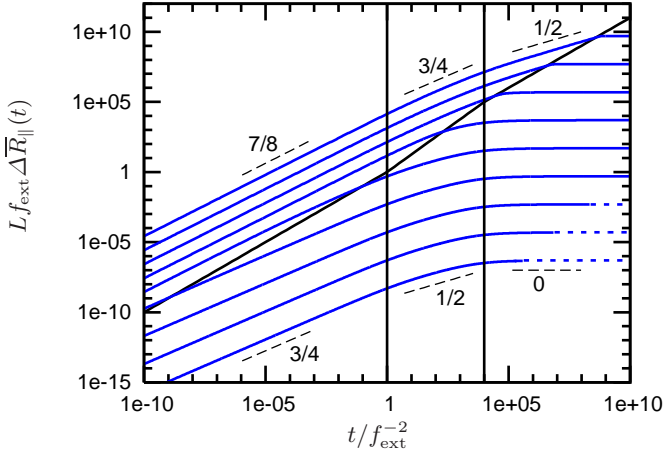}}}
  \caption{\label{fig:fppulling-deltar-C1em4}Numerical results for
    $\Delta \overline R_\parallel(t)$ for $\fpre/\fext=10^{-4}$ and
    $\Lambda^2=10^{-8\ldots8}$ from top to bottom. The dark vertical
    lines mark the crossovers at $t=\fext^{-2}$ (left) and
    $t=(\fext\fpre)^{-1}$ (right), while the dark diagonal line
    denotes the crossover at $t=\tlpar\simeq t_\star$, see
    table~\ref{tb:crossover-times}. The dashed lines indicate the
    asymptotic power laws of figs.~\ref{fig:phase-diagram-propagation}
    and \ref{fig:phase-diagram-relaxation}.}
\end{figure}

\begin{figure}
  \psfrag{XLABEL}{$t/\fext^{-2}$} \psfrag{YLABEL}{$-L\fext\Delta
    \overline R_\parallel(t)$} \centerline{\rotatebox{270}{
      \includegraphics{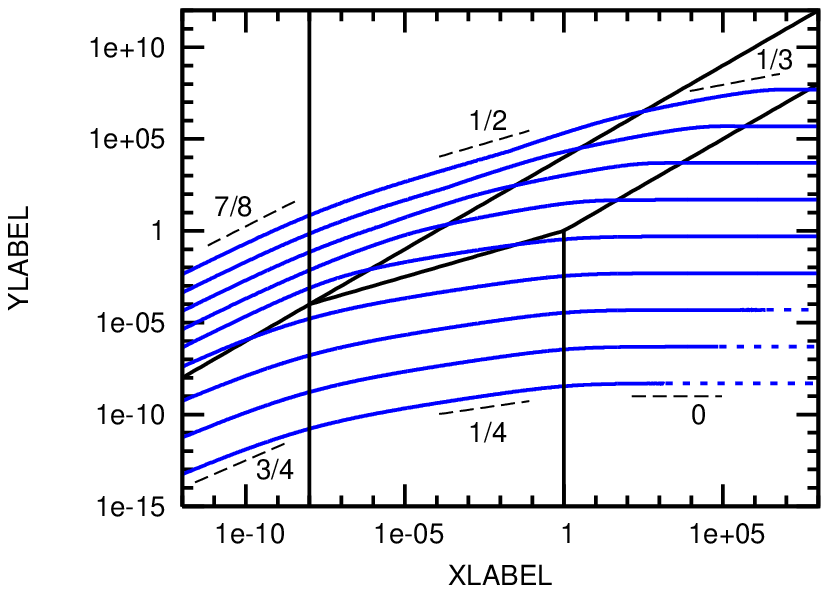}}}
  \caption{\label{fig:fppulling-deltar-C1e4}Numerical results for
    $\Delta \overline R_\parallel(t)$ for $\fpre/\fext=10^{4}$ and
    $\Lambda^2=10^{-8\ldots8}$ from top to bottom. The dark vertical
    lines mark the crossover at $t=\fpre^{-2}$ (left) and
    $t=\fext^{-2}$ (right), the dark diagonals mark the crossover at
    $t=\tlpar$ (upper) and $t=t_\star$ (lower). See
    table~\ref{tb:crossover-times}. The dashed lines indicate the
    asymptotic power laws of figs.~\ref{fig:phase-diagram-propagation}
    and \ref{fig:phase-diagram-relaxation}, and of
    eq.~\eqref{eq:deltar-hr}.}
\end{figure}

Using the algorithm outlined in sec.~\ref{sec:numerics}, we solved
eq.~\eqref{eq:pide-numerical} assuming a polymer of finite length $L$
for various values of the two parameters $c$ and $\Lambda^2$. The
observable $\Delta \overline R_\parallel(t)$ is then extracted from
the slope of the profile at $s=0$ and $s=L$ via eq.~\eqref{eq:deltar}.
As we analyzed the effect of the \emph{ratio} $\fpre/\fext$ of the two
force scales present in our scenario in the previous
sec.~\ref{sec:analytical-results}, we concentrate now on the
\emph{magnitude} of the force $\fext$ relative to the critical force
$\fc$ via the control parameter $\Lambda^2=\sqrt{\fc/\fext}$ for fixed
ratio $\fpre/\fext$.  In fig.~\ref{fig:fppulling-deltar-C1em4}, we
depict results for $\fpre/\fext=10^{-4}$ and various values of
$\Lambda^2$, i.e., for a sudden increase in stretching force.  One can
discern the intermediate regime at times $\fext^{-2}\ll t\ll
(\fext\fpre)^{-1}$, which vanishes if the stretching force is changed
only slightly. In the limit $\fpre=0$, it connects to the ``nonlinear
MSPT'' propagation ($t\ll\tlpar$) and ``nonlinear OPT'' relaxation
($t\gg t_\star$) regimes of the ``pulling''-scenario discussed in
ref.~\cite{hallatschek-frey-kroy:05}, respectively.
Fig.~\ref{fig:fppulling-deltar-C1e4} shows results for the inverse
scenario, a sudden decrease in the stretching force, with
$\fpre/\fext=10^4$.  Here, the time window $\fpre^{-2}\ll t\ll
\fext^{-2}$ produces a separate asymptotic regime only in the
relaxation phase which connects to the regime of nonlinear OPT
relaxation~\cite{hallatschek-frey-kroy:05} of the ``release''-scenario
if $\fext=0$. The time window $\tlpar\ll t\ll t_\star$ for the regime
of homogeneous tension relaxation with its $t^{1/3}$-scaling gets
broader as $\fpre/\fext\to\infty$.

Because of purely numerical noise in
figs.~\ref{fig:fppulling-deltar-C1em4} and
\ref{fig:fppulling-deltar-C1e4}, we have replaced the long-time
asymptote of $\Delta \overline R_\parallel(t)$ for $\Lambda^2\gg 1$
with dotted straight lines, because the subdominant
$\mathcal{O}(t^{-1/2})$-corrections in
eq.~\eqref{eq:deltar-premature-saturation} become smaller than the
numerical resolution and it is not possible anymore to reliably
evaluate eq.~\eqref{eq:deltar}.  The dashed lines in
figs.~\ref{fig:fppulling-deltar-C1em4} and
\ref{fig:fppulling-deltar-C1e4} indicate the asymptotic growth laws
for $\Delta \overline R_\parallel(t)$ in the respective regimes
summarized in figs.~\ref{fig:phase-diagram-propagation} and
\ref{fig:phase-diagram-relaxation}. By actually collapsing the
corresponding tension profiles onto the scaling functions derived in
sec.~\ref{sec:tension-profiles} and
appendix~\ref{app:analytics-solutions}, we find excellent quantitative
confirmation of our analytical approximations in the suitable
asymptotic limits (results not shown).

\subsection{Experimental implications}

The numerical results for $\Delta\overline R_\parallel(t)$ as shown in
figs.~\ref{fig:fppulling-deltar-C1em4} and
\ref{fig:fppulling-deltar-C1e4} over 20 time decades confirm the
intermediate asymptotics of eq.~\eqref{eq:pide}. They can also be used
to fit actual experiments, if one keeps in mind that these unbounded
asymptotic growth laws arise from superponing the exponential
relaxation of (infinitely) many modes of wavelength $0 < q < \infty$
with relaxation times $\tau_q\simeq q^{-2}/(q^2+f)$, see the response
function $\chi_\perp(q;t,t')$ [eq.~\eqref{eq:chi-perp}].  A real
polymer provides mode cutoffs at $q_\text{max}\simeq \pi/a$ where $a$
is some microscopic length scale (say, the polymer thickness) and at
$q_\text{min}\simeq \pi/L$. The latter cutoff defines a time scale
$\tlperp$, which is the relaxation time of the longest bending mode:
\begin{equation}\label{eq:tlperp}
  \tlperp \simeq 
  \begin{cases}
    L^4,&\text{if } \fext \ll L^{-2},\\
    L^2\fext^{-1},&\text{if }\fext \gg L^{-2}.
  \end{cases}
\end{equation}
For times beyond $\tlperp$, the continuum limit $L\to\infty$, i.e.,
disregarding the proper boundary condition and the discreteness of the
mode spectrum, becomes invalid; the algebraic relaxation ends and is
followed by normal exponential relaxation.  Since the crossover times
$\tlpar$, $t_\star$, and $\tlperp$ depend quite strongly on
experimental control parameters such as stretching force, filament
length, and persistence length, it is nevertheless possible to cover a
broad window of stretching and relaxation dynamics ($t\ll\tlperp$) by
adjusting these parameters. Our numerical solutions can easily be
adapted to the specific experimental setup at hand and provide a
quantitative description without free parameters.

\section{\label{sec:conclusion}Conclusions}

We have analyzed the generic scenario of a semiflexible filament
prestretched with a force $\fpre$ that is suddenly changed to $\fext$.
Related setups involving only one force scale have been discussed
previously~\cite{seifert-wintz-nelson:96,ajdari-juelicher-maggs:97,brochard-buguin-degennes:99,everaers-juelicher-ajdari-maggs:99},
but with diverse results based on partially contradicting
approximations. Extending a systematic theory in terms of backbone
tension~\cite{hallatschek-frey-kroy:05} to include two force scales
allows to resolve these discrepancies. Based on this tension formalism
(for its detailed formulation see
ref.~\cite{hallatschek-frey-kroy:07a}), we derived equations of motion
and motivated an intuitive blob-picture in order to estimate growth
laws for a central quantity, the boundary layer size or tension
propagation length $\lpar(t)$, which is a measure for how far (or how
fast) longitudinal correlations spread along the filament. The scaling
laws [eq.~\eqref{eq:scaling-lpar}] for $\lpar(t)$ have been confirmed
by computing tension profiles as scaling solutions to asymptotic
differential equations. We find that, for times beyond the initial
linear regime, which is dominated by the relaxation of very stiff
bending modes, the prestretching force $\fpre$ has a significant
influence. If $\fpre=0$, slowly relaxing long-wavelength contributions
in the initial conformation resist equilibration and dominate the
relaxation process: the taut-string approximation applies. If $\fpre$
is finite, these modes are removed from the initial configuration.
Hence, the dynamics equilibrates locally, which allows for a
quasi-static approximation.  Only if $\fpre\ll\fext$ is very small,
there is a crossover from the former to the latter situation. The
limiting single-force cases $\fext=0$ and $\fpre=0$, respectively,
analyzed in detail in ref.~\cite{hallatschek-frey-kroy:07b}, were also
recovered. However, the associated limits were found to be highly
nontrivial, which is reflected quite subtly in the corresponding
equations of motion for the tension.  These equations follow
systematically from the general eq.~\eqref{eq:pide}, while the blob
picture provides a more intuitive understanding of the underlying
process and the complicated intermediate asymptotics it produces. We
presented a numerical algorithm to solve eq.~\eqref{eq:pide} for very
general initial and boundary conditions, derived new predictions for
experimentally relevant observables and checked them against numerical
solutions.  Finally, the applicability of our results to experiments
was discussed. Our findings for the scenario analyzed in this work
should be relevant for the viscoelastic response of more complex
structures such as biopolymer
networks~\cite{gardel-etal:06,mizuno-tardin-schmidt-mackintosh:07}.
Further, we argue that the above established dependence of the
relaxation process on the \emph{initial} conditions also generalizes
to other force protocols, such as (weakly) time-dependent external
forces, transverse forces, elongational flows, or scenarios involving
sudden temperature changes.

\begin{acknowledgement}
  We gratefully acknowledge financial support via the German Academic
  Exchange Program (DAAD) (OH), by the Deutsche
  For\-schungsgemeinschaft through grant no. Ha 5163/1 (OH) and SFB
  486 (BO, EF), of the German Excellence Initiative via the program
  ``Nanosystems Initiative Munich (NIM)'' (BO, EF), and through
  BayEFG (BO).

\end{acknowledgement}

\appendix

\section{\label{app:analytics}Analytical approach (details)}

\subsection{\label{app:analytics-asymptotes}Asymptotes of
  eq.~\eqref{eq:pide}}

Following the analysis of ref.~\cite{hallatschek-frey-kroy:07b}, we
note that the Fourier integrals in eq.~\eqref{eq:pide} are dominated
by wavenumbers near a $\qm$ where the exponents are of order $1$.
Explicitly, we find for given time $t$ and integrated tension $F$ the
asymptotes:
\begin{subequations}\label{eq:qm}
  \begin{empheq}[left={\qm \simeq\empheqlbrace}]{align}
    &\label{eq:qm-1}
    t^{-1/4},&&\text{ if } F^2/t \ll 1\\
    &\label{eq:qm-2} F^{-1/2},&&\text{ if } F^2/t \gg 1
  \end{empheq}
\end{subequations}
Observing that $\qm\simeq \lperp^{-1}$ scales like the inverse of the
equilibration length $\lperp$ [eq.~\eqref{eq:scaling-lperp}], we
conclude that the first asymptote comprises free (bending-dominated)
relaxation where in the response function [eq.~\eqref{eq:chi-perp}]
the tension contribution $q^2 F$ is subdominant, whereas the second
one corresponds to forced (tension-driven) relaxation with the
subdominant bending contribution $q^4 t$.

Secondly, we write eq.~\eqref{eq:pide} as
\begin{equation}\label{eq:pide-D-S}
  \pd_s^2 F = D - S.
\end{equation}
The two contributions creating curvature in the tension profile have a
direct physical interpretation.
\begin{subequations}\label{eq:pide-DS}
  \begin{equation}\label{eq:pide-D}
    D=\int_0^\infty\!\frac{\td q}{\pi\lp} \frac{1-\e^{-2 q^2[q^2 t + F(s,t)]}}
    {q^2 + \fpre},
  \end{equation}
  is of \emph{deterministic} origin since it expresses the relaxation
  of initially excited modes for a polymer equilibrated under the
  force $\fpre$. The second term
  \begin{equation}\label{eq:pide-S}
    S=\int_0^\infty\!\frac{\td q}{\pi\lp}2 q^2\int_0^t\!\td t'\,\e^{-2 q^2 [q^2 (t-t')+F(s,t)-F(s,t')]}
  \end{equation}
  is of \emph{stochastic} origin since it comprises the accumulated
  influence of thermal excitations from times $t'<t$ mediated with the
  response function $\chi_\perp$ [eq.~\eqref{eq:chi-perp}].
\end{subequations}
With the general distinction [eq.~\eqref{eq:qm}] in mind, we first
turn our attention to the term $D$. In the bending-dominated case
[eq.~\eqref{eq:qm-1}] with $F^2/t\ll 1$, we neglect the tension
contribution $q^2 F$ in the exponent and obtain
\begin{equation}
  D\approx D_1 = \int_0^\infty\!\frac{\td q}{\pi\lp} \frac{1-\e^{-2 q^4 t}}
  {q^2 + \fpre} \qquad\text{if }F^2/t\ll 1,
\end{equation}
which possesses, depending on the magnitude of $\fpre$, the two
asymptotes
\begin{subequations}\label{eq:D-1}
  \begin{align}
    \label{eq:D-1a}
    D_{1,a} &\approx \frac{2^{3/4}t^{1/4}}{\EGamma{1/4}\lp}
    && \text{if } \fpre^2 t \ll 1, \quad F^2/t\ll 1\\
    \label{eq:D-1b}
    D_{1,b} &\approx \frac{1}{2 \lp \fpre^{1/2}} && \text{if } \fpre^2
    t \gg 1, \quad F^2/t \ll 1.
  \end{align}
\end{subequations}
In the tension-driven regime $F^2/t\gg 1$, on the other hand, we get
for $D$ by neglecting the bending contribution $q^4 t$
\begin{equation}
  D\approx D_2 = \int_0^\infty\!\frac{\td q}{\pi\lp} \frac{1-\e^{-2 q^2 F}}
  {q^2 + \fpre} \qquad\text{if } F^2/t\gg 1,
\end{equation}
which can be expanded for asymptotically small or large ratios of the
product $F \fpre$.  The leading-order terms are
\begin{subequations}\label{eq:D-2}
  \begin{align}
    \label{eq:D-2a}
    D_{2,a} &\approx \sqrt{\frac{2 F}{\pi\lp^2}}
    && \text{if } F \fpre \ll 1, \quad F^2/t\gg 1\\
    \label{eq:D-2b}
    D_{2,b} &\approx \frac{1}{2 \lp \fpre^{1/2}} && \text{if } F \fpre
    \gg 1, \quad F^2/t \gg 1.
  \end{align}
\end{subequations}

The stochastic term $S$ is simpler. In the regime $F^2/t \ll 1$ we
neglect the tension terms and simplify
\begin{equation}\label{eq:S-1}
  \begin{split}
    S\approx S_1&= \int_0^\infty\!\frac{\td q}{\pi\lp} 2 q^2
    \int_0^t\!\td
    t'\,\e^{-2 q^4 (t-t')} \\
    &= \int_0^\infty\!\frac{\td q}{\pi\lp} \frac{1-\e^{-2 q^4 t}}{q^2} \\
    &= \frac{2^{3/4} t^{1/4}}{\EGamma{1/4}\lp}, \qquad \text{if }
    F^2/t\ll 1.
  \end{split}
\end{equation}
For $F^2/t\gg 1$, the $t'$-integral in the stochastic term $S$ is
dominated by contributions near the upper limit and we can linearize
$F(s,t)-F(s,t')\approx \pd_t F(s,t) (t-t')$:
\begin{equation}\label{eq:S-2}
  \begin{split}
    S\approx S_2 &= \int_0^\infty\!\frac{\td q}{\pi\lp} 2 q^2
    \int_0^t\!\td t'\,
    \e^{-2 q^2(t-t')[q^2 + \pd_t F]} \\
    &= \int_0^\infty\!\frac{\td q}{\pi\lp} \frac{1-\e^{-2 q^2 t [q^2 +
        \pd_t F]}}
    {q^2 + \pd_t F} \\
    &\approx \int_0^\infty\!\frac{\td q}{\pi\lp} \frac{1}{q^2 + \pd_t F} \\
    &= \frac{1}{2\lp\sqrt{\pd_t F}},\qquad\text{if }F^2/t \gg 1.
  \end{split}
\end{equation}
In the third line we neglected the exponential in the limit $F^2/t\gg
1$, which expresses the instantaneous equilibration underlying the
quasi-static approximation of ref.~\cite{brochard-buguin-degennes:99};
this is an expansion similar to the one leading to
eq.~\eqref{eq:D-2b}.

\subsection{\label{app:analytics-solutions}Aymptotic equations of
  motion and solutions for the relaxation regime}

The asymptotic forms of eq.~\eqref{eq:pide} used in
sec.~\ref{sec:tension-profiles} were motivated physically and based on
approximations employed
previously~\cite{everaers-juelicher-ajdari-maggs:99,seifert-wintz-nelson:96,brochard-buguin-degennes:99},
but can also systematically be derived. The right hand side $D-S$ of
eq.~\eqref{eq:pide-D-S} was analyzed in
app.~\ref{app:analytics-asymptotes} for asymptotes of the
deterministic and stochastic term $D$ and $S$, respectively, see
eqs.~(\ref{eq:D-1},\ref{eq:D-2},\ref{eq:S-1},\ref{eq:S-2}).  Now we
can treat these various limits separately and find solutions for the
relaxation regimes omitted in sec.~\ref{sec:tension-profiles}. For
$t\gg t_\star$ we can exploit the fact that in this time regime the
tension deviates only by a small correction $\delta F$ from the flat
equilibrium profile $\fext t$. It is therefore possible to expand the
right hand side of eq.~\eqref{eq:pide} about this constant, which to
lowest order gives $\pd_s^2 \delta F \approx \left[D-S\right]_{F\to
  \fext t}$.

\begin{enumerate}
\item If $F^2/t\ll 1$ and $\fpre^2 t\ll 1$, we obtain $D\approx
  D_{1,a}$ and $S\approx S_1$, which cancel to leading order. But
  since the preconditions imply small $F$ and $\fpre$, the
  right-hand-side of eq.~\eqref{eq:pide-D-S} can be linearized with
  respect to these quantities. This was performed in
  sec.~\ref{sec:linear-regime}, see eq.~\eqref{eq:pide-linearized},
  and lead via a Laplace transform to the tension profile
  eq.~\eqref{eq:laplace-solution}. In the relaxation regime
  $L\ll\lambda$ with $\lambda(z)=2^{3/8}\lp^{1/2}z^{-1/8}$, we obtain
  the following tension profile:
  \begin{equation}
    \tilde F(s,z)=\frac{\fext}{z^2}+\frac{\Delta f}{z^2}\frac{s(s-L)}
    {2\lambda^2}.
  \end{equation}
  Backtransforming, we get
  \begin{equation}
    F(s,t) = \fext t + \frac{\Delta f\,t^{3/4}}{2^{7/4}\EGamma{7/4} \lp} s(s-L),
  \end{equation}
  i.e., a small correction $\delta F$ to the equilibrium tension
  $\fext t$ if $t\gg t_\star=L^8/\lp^4$.
\item $F^2/t \ll 1$ and $\fpre^2 t\gg 1$ gives $D\approx D_{1,b}$ and
  $S\approx S_1$, hence
  \begin{equation}
    \pd_s^2 F = D_{1,b}-S_1 \approx -S_1.
  \end{equation}
  This implies constant curvature (no propagation) in the time regime
  $\fpre^{-2}\ll t\ll \fext^{-2}$. The tension reads
  \begin{equation}
    F(s,t) = \fext t+\frac{2^{-1/4} t^{1/4} s(s-L)}{\EGamma{1/4}\lp}.
  \end{equation}
\item $F^2/t\gg 1$ and $F\fpre\ll 1$ yields
  \begin{equation}
    \pd_s^2 F \approx D_{2,a}-S_2 \approx D_{2,a},
  \end{equation}
  which comprises the taut-string case [eq.~\eqref{eq:ode-swn-1}]. We
  discussed the propagation regime in sec.~\eqref{sec:taut-string};
  the relaxation regime follows by writing $F(s,t)=\fext t+\delta
  F(s,t)$ and expanding, which yields
  \begin{equation}
    \delta F (s,t) = \frac{(\fext t)^{1/2}}{\sqrt{8\pi} \lp} s(s-L).
  \end{equation}
\item $F^2/t\gg 1$ and $F\fpre\gg 1$ leads to
  \begin{equation}
    \pd_s^2 F \approx D_{2,b} -S_2,
  \end{equation}
  which corresponds to the quasi-static approximation
  [eq.~\eqref{eq:ode-bbg}].  We discussed the propagation regime
  $t\ll\tlpar$ and the regime of premature tension saturation $t\gg
  t_\star$ in sec.~\ref{sec:quasi-static}. In the latter case, we used
  for eq.~\eqref{eq:deltaF-premature-saturation-improved} the
  next-to-leading order terms of the asymptotic expansions
  eqs.~(\ref{eq:D-2b},\ref{eq:S-2}). For $\fpre\gg\fext$ and $\fpre\gg
  \fc$ remains an intermediate regime $\tlpar\ll t\ll t_\star$ of
  homogeneous tension relaxation, see table~\ref{tb:crossover-times}.
  It turns out to be very related to the ``release''-case $\fext=0$ of
  ref.~\cite{hallatschek-frey-kroy:07b}. It is not possible to expand
  $F(s,t)=\fext\,t+\delta F$, since the correction $\delta F$ is for
  $t\ll t_\star$ not (yet) small compared to $\fext\,t$. However, as
  in ref.~\cite{hallatschek-frey-kroy:07b} for $\fext=0$, we can solve
  eq.~\eqref{eq:ode-bbg-derivative} using the separation ansatz
  $f(s,t)=g(t) h(s/L)$ and compute correction terms in the limit
  $\fpre\gg\fext$. Disregarding the proper initial conditions and
  assuming Dirichlet boundary conditions for the moment, we find that
  asymptotically
  \begin{equation}
    g(t) \sim \left(\frac{L^2}{\lp t}\right)^{2/3}.
  \end{equation}
  Although this is a decreasing function of time, we find that if
  $t\ll L^8/\lp^4$ the preconditions ($f^2 t\gg 1$ and $F\fpre\gg 1$)
  that lead to the asymptotic eq.~\eqref{eq:ode-bbg-derivative} are
  fulfilled.  In order to correct the Dirichlet boundary conditions
  assumed above to $h(0)=h(1)=\gamma(t)\equiv\fext/g(t)$, we require
  that $t\ll L^2/(\lp\fext^{3/2})$ which makes $\gamma\ll 1$. Both
  conditions reduce to $t\ll t_\star$,
  cf.~table~\ref{tb:crossover-times}. The roughly parabolic profile
  $h(\sigma)$ is a solution of
  \begin{equation}
    h''=- \frac{1}{6 h^{1/2}},\quad\text{with } h(0)=h(1)=\gamma(t),
  \end{equation}
  where a subdominant time-dependence stems from the small correction
  term $\gamma(t)$ to the Dirichlet boundary conditions.  It has the
  following characteristics:
  \begin{subequations}\label{eq:h-characteristics}
    \begin{align}
      \label{eq:h-characteristics-bulk}
      h(1/2) &= \frac{1}{16} \left(\frac{3}{2}\right)^{2/3}
      +\frac{1}{2}\gamma +    \mathcal{O}(\gamma^{3/2}), \\
      \label{eq:h-characteristics-slope}
      h'(0) &= 12^{-1/3} - \left(\tfrac{2}{3}\right)^{2/3}\gamma^{1/2}
      + \mathcal{O}(\gamma^{3/2}).
    \end{align}
  \end{subequations}

\end{enumerate}

\section{\label{app:numerics}Numerical approach (details)}

\subsection{\label{app:nonlinear-term}Evaluation of the nonlinear
  term}

The computation of the nonlinear term $G$ with its dependence on
$\Phi(\sigma,\tau_k)$ requires to evaluate the $z$-integral in
eq.~\eqref{eq:pide-numerical}. This can be done \emph{analytically},
if a piecewise linear interpolation is used in the exponent:
\begin{equation}\label{eq:piecewise-linearization}
  \Phi(\sigma,\tau_n)-z \Phi(\sigma,z \tau_n) \approx 
  A_k + B_k (z_{k+1}-z),
\end{equation}
for an index $k\in [0,n-1]$ such that $\tau_k\leq \tau_n z \leq
\tau_{k+1}$ and with $z_k=\tau_k/\tau_n$. The linearization
coefficients are
\begin{equation}\label{eq:linearization-coefficients}
  \begin{split}
    A_k &= \Phi(\sigma,\tau_n)
    - z_{k+1}\Phi(\sigma,\tau_{k+1})\\
    B_k &= \Phi(\sigma,\tau_{k+1}) + z_{k+1}
    \frac{\Phi(\sigma,\tau_{k+1})-\Phi(\sigma,\tau_k)}{z_{k+1}-z_k}.
  \end{split}
\end{equation}
With the linearization~\eqref{eq:piecewise-linearization},
$G[\Phi(\sigma,\tau_n)]$ is expressed as:
\begin{multline}\label{eq:G-linearized}
  G[\Phi(\sigma,\tau_n)] = \int_0^\infty\!\frac{\td \tilde q}{\pi}
  \Bigg[\frac{1-\e^{-2 \tilde q^2 \tau_n[\tilde q^2 +
      \Phi(\sigma,\tau_n)]}}
  {\tau_n(\tilde q^2 + c)}\Bigg] \\
  + \sum_{k=0}^{n-1}\int_0^\infty\!\frac{\td \tilde q}{\pi}\Bigg[
  \e^{-2 \tilde q^2 \tau_n[\tilde q^2 (1-z_{k+1}) + A_k]} \\
  \times \frac{\e^{-2 \tilde q^2 \tau_n (z_{k+1}-z_k)[\tilde q^2 +
      B_k]}-1} {\tau_n(\tilde q^2 + B_k)}\Bigg].
\end{multline}

It turns out that the nonlinear problems [eq.~\eqref{eq:bvp}] are
ill-conditioned in many cases. Their solutions are strongly affected
by numerical errors and often the algorithms used to solve these
problems fail to converge. Hence, it is necessary to evaluate the
nonlinear term [eq.~\eqref{eq:G-linearized}] as accurately as
possible. In order to represent the intermediate asymptotic regimes,
the variables $\tau$ and $\Phi$ have to vary over many orders of
magnitude. Thus, in eq.~\eqref{eq:G-linearized}, we have to take care
of overflow/underflow artefacts from the exponential functions. For
the numerical evaluation of the $\tilde q$-integrals, we need to make
sure that the dominant contribution from modes near $\qm=\tilde q_\text{m}\sqrt{\fext}$ is properly taken into account; see
eq.~\eqref{eq:qm}.

\subsection{\label{app:nlin-bvp}Nonlinear boundary value problems}

When solving a two-point boundary value problem with general mixed
linear boundary conditions such as eq.~\eqref{eq:bvp} (where we omit
the dependence on $\tau_n$), the problem is mapped to a system of
nonlinear equations. Writing the values $\Phi(\sigma_m)\equiv\Phi_m$,
where $m=0,\ldots, M$, as vector $\bvec
\Phi=(\Phi_0,\ldots,\Phi_M)^T$, the problem can be posed as
\begin{equation}\label{eq:nlin-bvp}
  \bvec T (\bvec \Phi)=0.
\end{equation}
The nonlinear operator $\bvec T$ is defined via the \emph{residuals}
of eq.~\eqref{eq:bvp}:
\begin{align}
  (\bvec T(\bvec \Phi))_m &= \Delta_m \bvec \Phi - G[\Phi_m],
  \qquad m=1,\ldots, M-1\nonumber\\
  (\bvec T(\bvec \Phi))_0 &= \alpha_{00}\Phi_0 + \alpha_{01}D_0^{+}
  \bvec\Phi
  - \alpha_{02}\label{eq:def-T} \\
  (\bvec T(\bvec \Phi))_M &= \alpha_{10}\Phi_M +
  \alpha_{11}D_M^{-}\bvec \Phi - \alpha_{12}\nonumber
\end{align}
Here, we have used discrete representations of the differential
operators $\pd_\sigma$ and $\pd_\sigma^2$. For the possibly
non-uniformly discretized coordinate $\sigma_m$ with local stepsize
$\hr=\sigma_{m+1}-\sigma_m$ and $\hl=\sigma_m-\sigma_{m-1}$, they are
given by:
\begin{equation}\label{eq:discrete-operators}
  \begin{split}
    \Delta_m \bvec \Phi &= \frac{2}{\hl (\hl+\hr)} \Phi_{m-1}
    - \frac{2}{\hl\hr}\Phi_m\\
    &\qquad + \frac{2}{\hr(\hl+\hr)}\Phi_{m+1} \\
    D_0^{+}\bvec \Phi &= \frac{1}{h_0^{+}}(\Phi_1-\Phi_0) \\
    D_M^{-}\bvec \Phi &= \frac{1}{h_M^{-}}(\Phi_M-\Phi_{M-1})
  \end{split}
\end{equation}

In the formulation [eq.~\eqref{eq:nlin-bvp}] our task is to find the
zero of $\bvec T$ in the vector space of discrete function
representations $\bvec\Phi$.  The common strategies are all based on
Newton's method~\cite{dennis-schnabel:83}. Given an initial guess
$\bvec\Phi_0$, this iterative method seeks a correction $\bvec d$ such
that $\bvec \Phi_0+\bvec d$ is the solution. If the initial guess is
good, the correction is small and we can write
\begin{equation}\label{eq:newt0}
  \bvec T(\bvec \Phi_0+\bvec d) \approx \bvec T(\bvec \Phi_0) 
  + J(\bvec \Phi_0)\bvec d=0
\end{equation}
with the Jacobian matrix
\begin{equation}\label{eq:jacobian}
  J_{ij}(\bvec\Phi) = \pdf{\bvec T(\bvec \Phi)_i}{\bvec \Phi_j}.
\end{equation}
This matrix can be computed either analytically or by
finite-difference approximation. In principle, the linear equation
[eq.~\eqref{eq:newt0}] can be solved straightforwardly for the optimal
correction $\bvec d$:
\begin{equation}\label{eq:newt0-solve}
  \bvec d = -J^{-1}(\bvec\Phi_0) \bvec T(\bvec \Phi_0).
\end{equation}
Finally, by iterating
\begin{equation}\label{eq:newtit}
  \bvec\Phi_{k+1}=\bvec\Phi_k-J^{-1}(\bvec\Phi_k)\bvec T(\bvec\Phi_k),
\end{equation}
one should finally find the solution $\bvec\Phi^\star =
\lim_{k\to\infty}\bvec\Phi_k$ with $\bvec T(\bvec \Phi^\star)=0$. For
``well-behaved'' operators $\bvec T$, this iteration converges
quadratically in the vicinity of the solution.  Unfortunately, in our
case $\bvec T$ is often ill-conditioned near the solution. More
sophisticated methods therefore modify the iteration
[eq.~\eqref{eq:newtit}]%
~\cite{dennis-schnabel:83}. Still, a very accurate evaluation of the
nonlinear term $G[\Phi]$ in eq.~\eqref{eq:def-T} is crucial to ensure
convergence. For our implementation, two different algorithms in the
GNU scientific library\footnote{The GSL is available from
  \texttt{www.gnu.org/software/gsl} and uses Powell's hybrid
  algorithm~\cite{powell:70a}.} and in the NLEQ package\footnote{The
  NLEQ package is available from \texttt{www.zib.de} and documented in
  ref.~\cite{nowak-weimann:91}.} have been used.

\subsection{\label{app:error-control}Error control}

We can control discretization errors stemming from representing the
continuous problem [eq.~\eqref{eq:pide-numerical}] as system of
nonlinear equations [eq.~\eqref{eq:nlin-bvp}] by reliable error
estimates.

An estimate for the spatial discretization error can be obtained via a
deferred corrections approach~\cite{pereyra:66}. The nonlinear system
[eq.~\eqref{eq:nlin-bvp}] is written with the low-order accurate
representation [eq.~\eqref{eq:discrete-operators}] for the spatial
derivatives.  Given its low-order accurate solution $\bvec\Phi$ and
another, high-order accurate operator $\bvec T^*$, we can solve the
perturbed system $\bvec T(\bvec \Phi^*)=-\bvec T^*(\bvec \Phi)$ to
obtain a high-order accurate solution $\bvec\Phi^*$. The
discretization error of $\bvec \Phi$ is estimated by the difference
$\bvec\Phi^*-\bvec\Phi$. The important point is that not the (possibly
complicated) high-order system associated with $\bvec T^*$, but only
the perturbed low-order system needs to be solved. A high-order
accurate operator $\bvec T^*$ is obtained by using more accurate
representations of the derivatives than those in
eq.~\eqref{eq:discrete-operators}, in our case by means of Noumerov's
method. Altogether, we obtain at least three digits of accuracy for
our choice of discretization.  We use ``graded'' grids,
$\sigma_m=\sigma_M (m/M)^\beta$ for $m=0,\ldots,M-1$.  For the
semiinfinite polymers, we take $\beta=1.25$ and dynamically adpat
$\sigma_M$; otherwise we take $\beta=1$ and $\sigma_M=1$.

The discretization error stemming from the $\tau$-discre\-ti\-za\-tion
needs to be estimated because we want to cover many orders of
magnitude in the time variable. Our results show that an exponential
growth in $\tau_n$ does not increase the discretization error (this is
due to the right hand side of eq.~\eqref{eq:pide} becoming
asymptotically independent of shorttime effects in the nonlinear
regime, cf.~eq.~\eqref{eq:S-2}). Hence, we choose $\tau_n=\Delta
(\delta^n-1)/(\delta-1)$ for $n=0,\ldots, N$ with the initial stepsize
$\Delta$ and the growth factor $\delta > 1$. To estimate the
discretization error, we take ``fine'' steps
$\tilde\tau_n=\tau_0(\delta^{n/2}-1)/(\delta-1)$ with $n=0,\ldots, 2N$
such that $\tilde\tau_{2n}=\tau_n$. In the step $n$ two additional
``fine'' solutions $\tilde\Phi(\sigma,\tilde\tau_{2n-1})$ and
$\tilde\Phi(\sigma,\tilde\tau_{2n})$ are computed. The error estimate
follows from the difference
$\tilde\Phi(\sigma,\tilde\tau_{2n})-\Phi(\sigma,\tau_n)$.  Altogether,
we get at least three digits of accuracy for our choice of the crucial
parameter $\delta$: for the semiinfinite polymers, we take $\delta =
1.1$ (only tension propagation), otherwise $\delta=1.05$ (also tension
relaxation).

\bibliography{journals,sf}

\end{document}